\documentclass[runningheads]{llncs}

\usepackage{eccv}

\usepackage{eccvabbrv}

\usepackage{graphicx}
\usepackage{booktabs}
\usepackage{multirow}
\usepackage{listings}
\usepackage{authblk}

\usepackage[a4paper, left=25.4mm, right=25.4mm, top=25.4mm, bottom=25.4mm]{geometry}

\newcommand{\ours}{\textsc{Madeleine}}
\newcommand{\oursSE}{\textsc{Madeleine-SE}}

\newcommand{\real}{\mathbb{R}}

\newcommand{\h}{\mathbf{h}}
\newcommand{\he}{\text{HE}}

\usepackage[accsupp]{axessibility}  

\usepackage{color} 
\definecolor{codegreen}{rgb}{0,0.6,0}
\definecolor{codegray}{rgb}{0.5,0.5,0.5}
\definecolor{codepurple}{rgb}{0.58,0,0.82}
\definecolor{backcolour}{rgb}{0.95,0.95,0.92}

\lstdefinestyle{mystyle}{
    backgroundcolor=\color{backcolour},   
    commentstyle=\color{codegreen},
    keywordstyle=\color{magenta},
    numberstyle=\tiny\color{codegray},
    stringstyle=\color{codepurple},
    basicstyle=\ttfamily\tiny,
    breakatwhitespace=false,         
    breaklines=true,                 
    captionpos=b,                    
    keepspaces=true,        
    lineskip=-1pt,
    numbers=left,                    
    numbersep=5pt,                  
    showspaces=false,                
    showstringspaces=false,
    showtabs=false,                  
    tabsize=2
}

\usepackage[pagebackref,breaklinks,colorlinks,citecolor=eccvblue]{hyperref}
\usepackage{hyperref}

\usepackage{orcidlink}

\begin{document}

\title{Multistain Pretraining for \\Slide Representation Learning in Pathology} 

\author{Guillaume Jaume\inst{1,2}$^,$\thanks{Co-first authorship.} \and
Anurag Vaidya\inst{1,2}$^{,\star}$ \and
Andrew Zhang\inst{1,2}$^,$\thanks{Co-second authorship.} \and
Andrew H. Song\inst{1,2}$^{,\star\star}$ \and \\ \vspace{-2.5mm}
Richard J. Chen\inst{1,2} \and
Sharifa Sahai\inst{1,2}\and
Dandan Mo\inst{2} \and
Emilio Madrigal\inst{2} \and \\ \vspace{-2.5mm} 
Long Phi Le\inst{1,2} \and 
Faisal Mahmood\inst{1,2}}

\authorrunning{G.~Jaume et al.}

\institute{
Harvard Medical School, Boston, MA, USA\\
\and
Mass General Brigham, Boston, MA, USA\\
\email{{gjaume,avaidya,faisalmahmood}@bwh.harvard.edu}
}

\maketitle

\begin{abstract}

Developing self-supervised learning (SSL) models that can learn universal and transferable representations of H\&E gigapixel whole-slide images (WSIs) is becoming increasingly valuable in computational pathology. These models hold the potential to advance critical tasks such as few-shot classification, slide retrieval, and patient stratification. Existing approaches for slide representation learning extend the principles of SSL from small images (e.g., 224$\times$224 patches) to entire slides, usually by aligning two different augmentations (or \emph{views}) of the slide. Yet the resulting representation remains constrained by the limited clinical and biological diversity of the \emph{views}. Instead, we postulate that slides stained with multiple markers, such as immunohistochemistry, can be used as different \emph{views} to form a rich task-agnostic training signal. To this end, we introduce $\ours$, a multimodal pretraining strategy for slide representation learning. $\ours$ is trained with a dual global-local cross-stain alignment objective on large cohorts of breast cancer samples (N=4,211 WSIs across five stains) and kidney transplant samples (N=12,070 WSIs across four stains). We demonstrate the quality of slide representations learned by $\ours$ on various downstream evaluations, ranging from morphological and molecular classification to prognostic prediction, comprising 21 tasks using 7,299 WSIs from multiple medical centers. Code is available at \href{https://github.com/mahmoodlab/MADELEINE}{https://github.com/mahmoodlab/MADELEINE}.

\end{abstract}

\section{Introduction}

Self-supervised learning (SSL) with multimodal pretraining is increasingly adopted in medical AI for constructing universal image representations that can be used for diagnosis, prognosis, and treatment response prediction~\cite{acosta2022multimodal,krishnan2022self,chen2024towards}. The core idea is to \textit{align} an image (\emph{e.g.,} a histology region-of-interest of a tumor) with other corresponding modalities (\emph{e.g.,} the morphological text description of the tumor) into a shared latent space via contrastive learning or other similarity matching losses~\cite{radford2021learning}. Intuitively, the richer the contrasting modality employed, the more detailed and nuanced the image representations can become, enabling better generalization and transferability to downstream tasks. 

In computational pathology~\cite{song2023artificial}, multimodal pretraining has mostly focused on building visual-language models of small images~\cite{gamper2021multiple,huang2023visual}, capitalizing on their success in computer vision~\cite{radford2021learning,wang2023image}. However, the scale of whole-slide images (WSIs), often exceeding 100,000 $\times$ 100,000 pixels at 20$\times$ magnification (0.5 $\mu$m/pixel), presents a significant challenge for adapting such techniques to pathology. To address this issue, most intra-modal and multimodal SSL methods focus on embedding small patches (\emph{e.g.,} 224 $\times$ 224-pixel images), which can then be aggregated using multiple instance learning (MIL) for downstream tasks~\cite{ilse2018attention,lu2021data,shao2021transmil}. Alternatively, the aggregation stage can also be pretrained with SSL to create a \emph{slide embedding} from the patch embeddings~\cite{tavolara2022contrastive,chen2022scaling,lazard2023giga,mukashyaka2024ebiomedicine}. The hierarchical construction from patches to patch embeddings to a slide embedding in a two-stage training pipeline enables self-supervised \emph{slide representation learning}, without utilizing labels from pathologists or learning task-specific representations.

However, most existing slide representation learning methods are intra-modal, thus limiting the richness and diversity of the training signal to learning visual invariances within the slide~\cite{chen2022scaling,lazard2023giga}. Instead, we propose to leverage additional modalities that naturally form clinically and biologically relevant \emph{pairs} suitable for pretraining. In this study, we hypothesize that WSIs stained with various markers, such as immunohistochemistry (IHC), can constitute a strong task-agnostic training signal for multimodal pretraining. Each stain can be seen as a different \textit{view} of the H\&E slide by highlighting spatially-resolved expression levels of relevant markers. In addition, unlike bulk gene expression data or text captions~\cite{jaume2024transcriptomics}, H\&E and other stains offer fine-grained morphological correspondences, which can be leveraged for enhanced representational power. 
 
To this end, we introduce $\ours$, an SSL approach for multistain-guided \emph{slide representation learning}. $\ours$ uses a multihead attention-based MIL~\cite{ilse2018attention,lu2021data} to encode pre-extracted patch embeddings into a \emph{slide embedding}. $\ours$ is pretrained on large collections of multistain tissue using a dual global-local cross-stain objective. The global objective, based on a symmetric contrastive loss~\cite{chen2020simple}, learns slide-level correspondences between the H\&E slide and the other stains. This alignment guides the H\&E embedding to encapsulate the global morphological composition of the tissue. The local objective, based on the Graph Optimal Transport framework~\cite{chen2020graph,pramanick2022volta}, learns patch-level correspondences between the H\&E and the other stains, thereby enabling cross-stain matching of fine-grained morphological features. The resulting latent space (i) can encode all stains encountered during pretraining, as the same network is employed for encoding each stain, and (ii) can be used for diverse downstream applications, as the training signal and resulting model are task-agnostic. 

To summarize, our contributions are
(1) $\ours$, a multimodal pretraining strategy for \emph{slide representation learning} in computational pathology;
(2) a large-scale demonstration of $\ours$ pretraining on two organs, breast (N=4,211 slides, five stains) and kidney (N=12,070 slides, four stains); and
(3) extensive evaluation of $\ours$ across 21 tasks including morphological subtyping, molecular subtyping, survival prediction, and IHC quantification, tested in various scenarios for few-shot learning (using linear probing and prototyping) and model fine-tuning.

\section{Related work}

\begin{figure*}[t]
   \centering
   \includegraphics[width=0.99\linewidth]{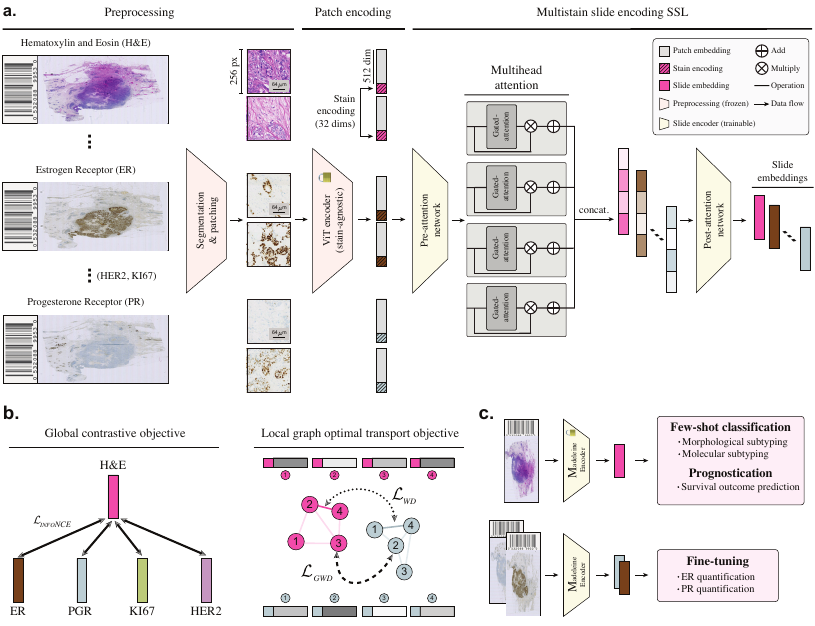}
   \caption{\textbf{Overview of $\ours$.}
   \textbf{a.}
   \textbf{Preprocessing:} WSIs from various stains undergo tissue segmentation and patching into 256$\times$256-pixel tiles.
   \textbf{Patch encoding:} All patches are passed through a stain-agnostic Vision Transformer to extract patch embeddings augmented with a learnable stain-specific encoding. 
   \textbf{Slide encoding:} Embeddings from each stain are sequentially passed through a pre-attention, a multi-head attention, and a post-attention module, resulting in stain-specific slide embeddings.
   \textbf{b.} $\ours$ is trained with a combination of global and local objectives.
   \textbf{Global objective:} Slide embeddings are aligned using a cross-modal contrastive objective (\textsc{infoNCE}).
   \textbf{Local objective:} Patch embeddings are aligned using a cross-modal local \textsc{Graph Optimal Transport} objective.
   \textbf{c.} The resulting stain-agnostic slide encoder can be used for various downstream tasks in few-shot and full fine-tuning settings. 
   }
   \label{fig:overview}
\end{figure*}

\subsection{Vision representation learning}

Training a Vision Transformer (ViTs)~\cite{vaswani2017attention,dosovitskiy2020image} with self-supervised learning (SSL) \cite{zhou2022image,caron2021emerging} is now the preferred approach for learning task-agnostic image representations, such as based on visual-language models~\cite{radford2021learning,jia2021scaling,li2021align,singh2022flava,yu2022coca,alayrac2022flamingo,li2023blip,li2023scaling,wang2023image}.
Visual-language models are usually based on contrastive learning~\cite{radford2021learning}, where the objective is to maximize the similarity between an image and its textual description, or as recently proposed, using Optimal Transport (OT) for fine-grained cross-modal alignment~\cite{kim2022differentiable,chen2020uniter,pramanick2022Ovolta}. This approach is framed as a distribution matching objective, where the aim is to minimize the cost associated with a transport plan to match a token distribution of one modality to the other. Differently, multimodal training can leverage other \emph{spatial} modalities, such as depth maps or bounding box annotations~\cite{bachmann2022multimae}. Drawing on these methodologies, our model, $\ours$, integrates various high-resolution ``views'' of the \emph{same} tissue stained with different markers, such as estrogen or progesterone receptor stainings.

\subsection{Representation learning of histology images}

SSL for learning representations of histology images is an active field with efforts in (i) developing models that can extract embeddings from small patches, typically 256$\times$256 in size, and (ii) creating models designed to derive representations from entire WSIs, a task we denote as \emph{slide representation learning}, and which constitutes the central contribution of our study.

\noindent\textbf{Patch representation learning} Using SSL to encode histology patches has so far been the main focus with increasingly large models trained on larger datasets~\cite{koohbanani2021self,filiot2023scaling,vorontsov2023virchow,kang2023benchmarking,wang2022transformer,azizi2023robust,chen2024towards,xu2024gigapath} (e.g., \cite{campanella2023computational} used 3 billion patches from 423,000 slides). Simultaneously, vision-language models for histopathology have been developed using large datasets from sources such as social media and educational textbooks~\cite{gamper2021multiple,huang2023visual,lu2024towards}. Similar to $\ours$, \cite{hua2023pathoduet} proposed multimodal fine-tuning by aligning H\&E and IHC patches. However, their method focuses on encoding patches, whereas $\ours$ focuses on encoding WSIs. 

\noindent\textbf{Slide representation learning} Developing pretrained encoders that extend beyond simple regions of interest to gigapixel whole slides is the next frontier in representation learning of histology images. Several works  \cite{chen2022scaling,lazard2023giga,yu2023slpd,jiang2023masked,aryal2023context,vu2023handcrafted,tavolara2022contrastive,mukashyaka2024ebiomedicine,song2024morphological,song2024multimodal} proposed hierarchical slide pretraining, first by transforming each patch into a patch embedding and then into a slide embedding (or region embedding).
The slide encoder is typically trained using image augmentation techniques to define different \textit{views} of the slide followed by contrastive or reconstruction objectives. 
Concurrent to this work, multimodal pretraining for slide representation learning was explored using bulk transcriptomics~\cite{jaume2024transcriptomics} and pathology reports~\cite{shaikovski2024prism}.

\subsection{Beyond H\&E staining}

While H\&E staining remains the gold standard in standard-of-care, it is often complemented with immunohistochemistry (IHC) and special stains. Several works have been proposed for automatic IHC quantification~\cite{vandenberghe2017relevance,khameneh2019automated,qaiser2018her2,ghahremani2022deep}, often leveraging cell segmentation networks. Differently, IHC status can be predicted from H\&E slides, such as for HER2 (human epidermal growth factor receptor 2) status prediction in invasive breast cancer or EGFR (epidermal growth factor receptor) prediction in lung cancer~\cite{wang2023hahnet,couture2018image,shamai2019artificial,naik2020deep,kather2020pancancer,anand2020deep,farahmand2022deep,akbarnejad2023predicting,shamai2022deep,rawat2020deep}.
Stain transfer, also known as virtual staining has also been proposed~\cite{rivenson2019virtual,lahiani2020seamless,zeng2022semi,bai2023deep,dehaan2021deep,liu2022bci,liu2021unpaired,boyd2022medical,lin2022unpaired}, in particular based on unpaired style transfer techniques~\cite{zhu2017unpaired,park2020contrastive,isola2016image}. 

\section{Methods}

We introduce $\ours$ for multistain-guided slide representation learning (Fig.~\ref{fig:overview}). $\ours$ is composed of (1) a stain-agnostic patch encoder that transforms histology patches into \emph{patch embeddings} (Sec.~\ref{sec:preprocessing} and \ref{sec:conch}), (2) a multihead attention-based MIL to learn a \emph{slide embedding} (Sec.~\ref{sec:abmil}), and (3) a cross-stain alignment module based on a dual global-local objective (Sec.~\ref{sec:alignment}).

\subsection{Pre-processing and notation} \label{sec:preprocessing}

Given a histology slide $\mathbf{X}_i \in \real^{d_x \times d_y \times 3}$ (H\&E or another stain) for the $i^{\text{th}}$ patient, we follow the MIL paradigm~\cite{ilse2018attention,lu2021data,shao2021transmil,lee2022derivation,li2021dual}, which consists of tessellating the slide into small patches, using a pretrained vision encoder to extract patch embeddings, and pooling the resulting patch embeddings into a slide embedding. We use $s_k$ to refer to the $k^{\text{th}}$ stain with $\{s_k \}_{k=1}^K$ collectively referring to all non-H\&E  stains, \emph{e.g.,} in breast cases, $s_k\in\{\text{ER}, \text{PR}, \text{HER2}, \text{KI67}\}$ with $K=4$ denoting estrogen receptor, progesterone receptor, human epidermal growth factor receptor 2, and antigen kiel 67, respectively. We start by detecting and segmenting tissue regions to discard any background information. We use the CLAM toolbox~\cite{lu2021data} to detect H\&E tissue and employ a deep learning-based tissue detector trained on mask annotations to detect non-H\&E tissue. We then extract non-overlapping 256$\times$256 patches on all stains. 

\subsection{Patch encoding} \label{sec:conch}

As $\ours$ is trained on multiple stains, this renders most SSL models for patch feature extraction trained on H\&E suboptimal~\cite{wang2021transpath,vorontsov2023virchow}. Instead, we use CONCH, the image encoder of a visual-language model pretrained on 1M histology image-caption pairs curated from existing publications, which includes various histology stains~\cite{lu2024towards}. We obtain the H\&E patch embeddings $\mathbf{H}_i^{\text{HE}}\in\mathbb{R}^{N_{\text{HE}}\times d}$, with $N_{\text{HE}}$ and $d=512$ denoting the number of H\&E patches and the embedding dimension, respectively. The $j^{\text{th}}$ row entry,  $\mathbf{H}_{i,j}^{\text{HE}}$, corresponds to the $j^{\text{th}}$ patch embedding. We perform the same procedure for other non-H\&E stains $\{s_k\}_{k=1}^K$ to obtain patch embeddings, i.e., $\mathbf{H}_i^{s_k}\in\mathbb{R}^{N_{s_k}\times d}$.

\subsection{Slide encoding} \label{sec:abmil}

\textbf{Pre-attention \& stain encoding} The patch embeddings $\mathbf{H}_{i}^{\text{HE}}$ are first passed through a \textit{pre-attention} network, $f^{\text{pre}}: \mathbb{R}^{d} \rightarrow \mathbb{R}^{d}$, resulting in $\widetilde{\mathbf{H}}_{i}^{\text{HE}}\in\mathbb{R}^{d\times d}$.
As the same pre-attention module is used for encoding \emph{all} stains, having a stain-specific signature in the input can be beneficial. To do so, we define a learnable stain-specific encoding (denoted as SE, 32 dims) that is concatenated to each patch token before pre-attention, with $d=d+32$. This is inspired by modality-specific token augmentation schemes in multimodal fusion~\cite{jaegle2022perceiver,liang2023highmodality,song2024multimodal}. 

\noindent\textbf{Multi-head attention-based MIL} We subsequently pass the resulting patch embeddings $\widetilde{\mathbf{H}}_{i}^{\text{HE}}$ to a multihead (MH) attention network with $M$ heads~\cite{ilse2018attention}, resulting in an attention score $a_{i,j}^m\in[0,1]$ for each patch (Appendix Equation 2).
Using multiple attention heads allows each head to focus on different yet morphologically important regions, similar to multi-head attention in Transformers~\cite{vaswani2017attention,dosovitskiy2020image}. Once computed, we form head-specific slide embeddings by taking the weighted average of the transformed patch embeddings, \emph{i.e.,} $\mathbf{h}^{\text{HE}}_{i,m}=\sum_{j=1}^{N_\he} a^m_{i,j}\widetilde{\mathbf{H}}_{i,j}^{\text{HE}}$. The resulting slide embedding $\mathbf{h}_i^{\he}$ is formed by concatenating the $M$ slide embeddings and passing it through a \textit{post-attention} network for dimension reduction, $f^{\text{post}}: \mathbb{R}^{Md}\rightarrow \mathbb{R}^{d}$,
\begin{equation}
    \mathbf{h}_i^{\he} = f^{\text{post}}([\h_{i,1}^{\he}, \ldots, \h_{i,M}^{\he}]).
\end{equation}
The slide embeddings for other stains $\{\h_{i}^{s_k}\}_{k=1}^K$ are computed analogously. We emphasize that we apply the \emph{same} model to all stains, instead of stain-specific modules. This way, we reduce memory requirements by a factor of $K$ (the number of stains) and constrain the network to learn stain-agnostic representations.

\subsection{Loss} \label{sec:alignment}

$\ours$ is trained using a combination of two cross-modal objectives: (1) a global objective to align slide embeddings of all stains in a shared latent space, and (2) a local objective for matching cross-stain patch embeddings. We optionally complement these two objectives with an intra-modal loss. 

\noindent\textbf{Cross-modal global alignment (\textsc{infoNCE})} We align the latent space induced by each stain through a global symmetric cross-modal contrastive learning objective, commonly referred to as \textsc{infoNCE}~\cite{chen2020simple}. This is a widely employed representation learning formulation~\cite{radford2021learning}, especially in visual-language pretraining. This objective enforces slide embeddings from the same case to be closer to each other while pushing away slide embeddings from different cases. Each term maximizes the dot-product similarity between embeddings from the same pair normalized (with Softmax) by negative pairs, which can be seen as other ``classes".

\noindent\textbf{Cross-modal local alignment (\textsc{GOT})} We also perform \emph{local} alignment by matching the empirical distributions of patch embeddings of all stains. Intuitively, as the local morphological structure is preserved across different stains, we can identify fine-grained cross-stain correspondences. The model can consequently learn to distinguish H\&E morphologies corresponding to marker-positive and marker-negative regions. 

To this end, we leverage the framework of graph optimal transport (GOT)~\cite{chen2020graph,pramanick2022volta}. Formally, we define the empirical distribution of the H\&E patch embeddings as $\hat{p}_{\text{HE}}=\frac{1}{N_{\text{HE}}}\sum_{j=1}^{N_{\text{HE}}}\delta(\mathbf{H}_{i,j}^{\text{HE}})$, with $\delta(\cdot)$ denoting the dirac-delta function. We additionally define an H\&E graph $\mathcal{G}_{\text{HE}}(V_{\text{HE}}, E_{\text{HE}})$, where the node $v^{\he}_j\in V_{\text{HE}}$ represents the $j^{\text{th}}$ patch embedding from $\mathbf{H}_{i,j}^{\text{HE}}$ and the edge $e^{\he}_{j,j'}$ is formed if the cosine similarity between $v^{\he}_j$ and $v^{\he}_{j'}$ is above a certain threshold.  The same construction is applied to all other stains. 

Based on this setup, we aim to cross-align the stain-specific graphs by minimizing two metrics: (i) The Wasserstein distance (WD) $\mathcal{L}_{\text{Node}}(\hat{p}_{\text{HE}}, \hat{p}_{s_k})$ defined between the empirical distributions of patch embeddings (\emph{i.e.,} the nodes of $\mathcal{G}_{\text{HE}}$ and $\mathcal{G}_{s_k}$). Intuitively, WD can be seen as computing the distance between the node embedding distributions of different stains. (ii) The Gromov-Wasserstein distance (GWD) $\mathcal{L}_{\text{Edge}}(\hat{p}_{\text{HE}}, \hat{p}_{s_k})$ between the edges of $\mathcal{G}_{\text{HE}}$ and $\mathcal{G}_{{s_k}}$. Intuitively, GWD enforces stain-specific graphs to follow a similar structure (or topology). Additional technical information is provided in Appendix 1.3.

The local alignment objective $\mathcal{L}_{\text{GOT}}$ is given as the combination of two metrics over cross-stain pairs, with $\gamma$ denoting a weighting term, 
\begin{equation}
    \mathcal{L}_{\text{GOT}} = \gamma \sum_{k=1}^K\mathcal{L}_{\text{Node}}(\hat{p}_{\text{HE}}, \hat{p}_{s_k}) + (1-\gamma)\sum_{k=1}^K\mathcal{L}_{\text{Edge}}(\hat{p}_{\text{HE}}, \hat{p}_{s_k}).
\end{equation}

\noindent\textbf{Intra-modal alignment (\textsc{Intra})} We additionally define an optional intra-modality objective $\mathcal{L}_{\textsc{Intra}}$ to align different augmentations of the H\&E slide. This objective is similar to existing pretraining strategy~\cite{chen2022scaling,lazard2023giga}, and can be seen as a direct extension of SSL from patch- to slide-level. Specifically, we generate two distinct slide embeddings of the H\&E slide by separately processing two randomly disjoint sets of patch embeddings using $\ours$. This process yields a pair of slide embeddings, denoted as $\h_i^{\he, \text{(1)}}$ and $\h_i^{\he, \text{(2)}}$, which are then aligned using a contrastive objective.

Overall, we train $\ours$ with the composite loss $\mathcal{L}=\mathcal{L}_{\text{InfoNCE}} + \mathcal{L}_{\text{GOT}}$. The \textsc{Intra} objective is used as a baseline, which can also be combined with $\mathcal{L}$.

\noindent\textbf{Pretraining details} $\ours$ was trained for a maximum of 120 epochs (5 warmup epochs) using AdamW optimizer, cosine learning rate schedule (start: $10^{-4}$, end: $10^{-8}$) and batch size of 90. All models were trained on 3$\times$24 GB 3090Ti. Additional implementation details are provided in Supplementary Table 1.

\section{Study design} \label{sec:predataset}

To assess the representative power of $\ours$ pretraining, we design two distinct scenarios: (1) $\ours$ pretraining on breast cancer cases (Sec.~\ref{sec:breast}) and (2) $\ours$ pretraining on kidney transplant cases (Sec.~\ref{sec:kidney}). We then perform downstream evaluations based on public and private cohorts (Sec. \ref{sec:eval}). The evaluation was designed to encompass the variability of tasks found in pathology. We emphasize that $\ours$ pretraining does not involve datasets used for downstream tasks, precluding any data leakage. A detailed description is provided in Supplementary 2.

\subsection{Breast} \label{sec:breast}

\noindent\textbf{Acrobat (multi-stain, pretraining)} We pretrain $\ours$ using data from the Automatic registration of breast cancer tissue MICCAI challenge (Acrobat)~\cite{weitz2022acrobat,weitz2023multi}. Acrobat is a multi-stain dataset comprising 4,211 WSIs from 1,153 primary breast cancer cases. Every case includes an H\&E-stained WSI, along with one to four WSIs of tissue from the same tumor that have been stained with immunohistochemistry, either ER (N=844 WSIs), PR (N=837), HER2 (N=534), or KI67 (N=843), such that K=4. The entirety of Acrobat was used for pretraining, with all slides processed at 10$\times$ magnification. 

\noindent\textbf{TCGA Breast (H\&E, downstream)} We use the public TCGA Breast cohort for (1) morphological subtyping (N=1,041) into invasive ductal carcinoma (IDC) and invasive lobular carcinoma (ILC); (2) binary molecular subtyping for predicting ER status (N=996), PR status (N=993), and HER2 status (N=693), and (3) survival prediction (N=1,049).

\noindent\textbf{BCNB (H\&E, downstream)} We use the public BCNB cohort~\cite{xu2021predicting} for binary molecular subtyping by predicting ER, PR, HER2 and KI67 status (N=1,058).

\noindent\textbf{AIDPATH (H\&E, downstream)} We use the public AIDPATH cohort~\cite{aidpath2024} for binary HER2 status prediction (N=48) and KI67 status prediction (N=50). 

\noindent\textbf{BWH Breast (H\&E, downstream)} We use an in-house breast cohort for two binary tasks (1) morphological subtyping (N=1,265); and (2) molecular subtyping for predicting ER (N=873), PR (N=874), and HER2 (N=816). 

\noindent\textbf{MGH Breast (ER and PR, downstream)} We use another in-house breast cohort for IHC quantification of ER abundance (N=962) and PR abundance (N=1,071). We frame both ER and PR quantification as 3- and 6-class problems.

\subsection{Kidney} \label{sec:kidney}

\noindent\textbf{BWH Kidney (multi-stain, pretraining)} We collected an in-house renal transplant cohort comprising kidney biopsies from 1,069 renal transplant cases. Each case includes one to three tissue blocks, where each block consists of one to two H\&E-stained ($N$=4,638) and one to two periodic acid-Schiff (PAS) ($N$=4,630) slides, one Jones-stained slide ($N$=2,326) and one Trichrome-stained slide ($N$=2,328), such that $K=3$. In total, each case includes 6 to 18 slides. We hold out 20\% of this cohort (210 cases, 1,852 slides across all stains, out of which 463 are H\&E slides) as an independent test set and used the rest for $\ours$ pretraining. Slides were processed at 20$\times$.

\noindent\textbf{Renal allograft rejection (downstream)} We use H\&E slides of the held-out cohort (N=463) to screen for Antibody-mediated rejection (AMR, 2 class) and quantify Interstitial Fibrosis and Tubular Atrophy (IFTA, 3 classes) (N=210 cases, N=1,852 WSIs across all stains). As each case includes several H\&E slides, we define two sub-tasks: ``single-slide'' prediction, where we use a single slide per case (N=463 H\&E slides), and ``all-slides'' prediction, where we use all available slides per case (N=305 H\&E slides). 

\subsection{Evaluation framework} \label{sec:eval}

\noindent\textbf{Few-shot classification} Following the standard practice in SSL evaluation~\cite{caron2021emerging,zhou2022image,chen2024towards}, we benchmark $\ours$ and baselines with $k$-shot classification ($k=1,5,10,25$ examples per class) using (1) linear probing and (2) prototyping. All experiments are repeated ten times by randomly sampling $k$ examples per class. Linear probing was conducted without hyper-parameter search using default parameters of the sklearn package. 

\noindent\textbf{Survival prediction} We assess $\ours$ in survival outcome prediction, where slide embeddings are passed to a Cox proportional hazards loss predicting survival. Following prior work, MIL models are trained using survival negative log-likelihood (NLL) loss~\cite{chen2022pan}. We use site- and survival-stratified five-fold cross-validation evaluated with concordance-index (c-index) \cite{jaume2023modeling}.  

\noindent\textbf{Fine-tuning} We assess the performance of $\ours$ encoder for downstream tasks when fine-tuned, compared to when trained from scratch. Evaluations follow a 5-fold label-stratified train-test strategy.

\section{Results}

\begin{figure*}[t]
   \centering
   \includegraphics[width=0.95\linewidth]{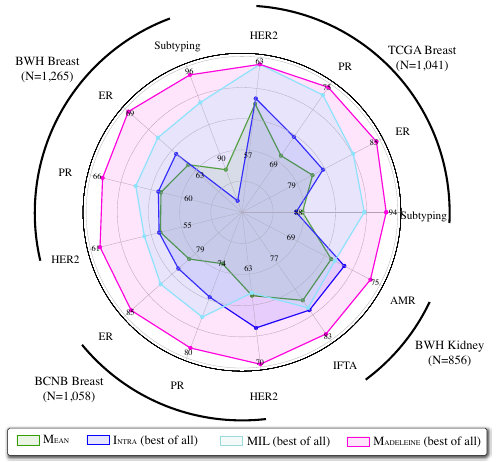}
   \caption{\textbf{Few-shot performance of $\ours$ against baselines.} All tasks are assessed on H\&E-stained WSIs. Morphological subtyping is reported for $k$=10, molecular subtyping for $k$=25, and kidney transplant rejection for $k$=50. Each experiment is repeated ten times by sampling $k$ different samples per class. Besides HIPT and GigaSSL, all models use the \emph{same} patch encoder. Each axis represents 10\% AUC and each segment a 2\% increment. Additional results for all $k$ values are provided in Supplementary 4 and 5.
   } 
   \label{fig:fig2}
\end{figure*}

\begin{table*}[t]
\centering
\caption{\textbf{Few-shot breast cancer subtyping.} Evaluation using macro-AUC on TCGA Breast (N=1,041) and BWH Breast cohorts (N=1,265). GigaSSL embeddings for BWH cohort not available. Mean and standard deviation reported over ten runs. Best in \textbf{bold}, second best is \underline{underlined}.}
\label{tab:brca_fewshot}
\scalebox{1.0}{
\begin{tabular}{ll|cccc|cccc}
\toprule
& Model/Data & \multicolumn{4}{c}{\textbf{TCGA Breast} ($\uparrow$)}  & \multicolumn{4}{c}{\textbf{BWH Breast} ($\uparrow$)}\\
& & $k$=1 & $k$=5 & $k$=10 & $k$=25 &  $k$=1 &  $k$=5 &  $k$=10 &  $k$=25 \\
\midrule

\parbox[t]{3mm}{\multirow{4}{*}{\rotatebox[origin=c]{90}{\textbf{MIL}}}} 
& \multirow{1}{*}{ABMIL \cite{ilse2018attention}} & 79.7 \scriptsize{$\pm$ 11.8} & 90.4 \scriptsize{$\pm$ 3.8} & 90.4 \scriptsize{$\pm$ 4.1} & 93.3 \scriptsize{$\pm$ 1.2} & 67.8 \scriptsize{$\pm$ 13.3} & 84.7 \scriptsize{$\pm$ 9.3} & 93.0 \scriptsize{$\pm$ 2.9} & 95.3 \scriptsize{$\pm$ 2.0} \\

& \multirow{1}{*}{TransMIL \cite{shao2021transmil}}  & 61.8 \scriptsize{$\pm$ 5.7} & 71.5 \scriptsize{$\pm$ 10.5} & 72.1 \scriptsize{$\pm$ 10.6} & 82.5 \scriptsize{$\pm$ 6.4} & 59.5 \scriptsize{$\pm$ 12.2} & 72.6 \scriptsize{$\pm$ 14.0} & 73.6 \scriptsize{$\pm$ 9.4} & 81.2 \scriptsize{$\pm$ 12.5} \\

& \multirow{1}{*}{IB-MIL \cite{li2023task}} & 74.8 \scriptsize{$\pm$ 10.8} & 90.1 \scriptsize{$\pm$ 4.5} & 91.4 \scriptsize{$\pm$ 2.2} & 93.8 \scriptsize{$\pm$ 1.5} & 68.7 \scriptsize{$\pm$ 13.1} & 82.6 \scriptsize{$\pm$ 8.0} & 88.7 \scriptsize{$\pm$ 6.3} & 95.2 \scriptsize{$\pm$ 1.9} \\

& \multirow{1}{*}{ILRA \cite{xiang2022exploring}}  & 68.3 \scriptsize{$\pm$ 9.3} & 89.3 \scriptsize{$\pm$ 2.9} & 90.9 \scriptsize{$\pm$ 1.6} & 92.2 \scriptsize{$\pm$ 2.1} & 69.8 \scriptsize{$\pm$ 8.6} & 84.7 \scriptsize{$\pm$ 7.6} & 91.0 \scriptsize{$\pm$ 2.4} & 93.3 \scriptsize{$\pm$ 3.3} \\

\midrule

\parbox[t]{3mm}{\multirow{7}{*}{\rotatebox[origin=c]{90}{\textbf{Linear probe}}}} 
& \multirow{1}{*}{\textsc{Mean}}  & 70.5 \scriptsize{$\pm$ 11.0} & 83.1 \scriptsize{$\pm$ 3.7} & 86.6 \scriptsize{$\pm$ 3.2} & 91.2 \scriptsize{$\pm$ 1.2} & 70.2 \scriptsize{$\pm$ 9.3} & 81.8 \scriptsize{$\pm$ 5.4} & 87.4 \scriptsize{$\pm$ 2.8} & 92.6 \scriptsize{$\pm$ 1.5} \\

& \multirow{1}{*}{\textsc{Intra}} & 70.8 \scriptsize{$\pm$ 9.1} & 82.6 \scriptsize{$\pm$ 4.2} & 86.1 \scriptsize{$\pm$ 2.9} & 91.2 \scriptsize{$\pm$ 1.2} & 69.3 \scriptsize{$\pm$ 8.7} & 78.3 \scriptsize{$\pm$ 7.1} & 84.8 \scriptsize{$\pm$ 3.0} & 92.0 \scriptsize{$\pm$ 2.4} \\

& \multirow{1}{*}{HIPT$_{\text{CLS-4k}}$ \cite{chen2022scaling}}  & 62.2 \scriptsize{$\pm$ 3.9} & 69.3 \scriptsize{$\pm$ 5.4} & 77.5 \scriptsize{$\pm$ 3.9} & 83.0 \scriptsize{$\pm$ 2.3} & 66.8 \scriptsize{$\pm$ 12.6} & 76.6 \scriptsize{$\pm$ 6.2} & 80.6 \scriptsize{$\pm$ 3.3} & 85.8 \scriptsize{$\pm$ 2.0} \\

& \multirow{1}{*}{GigaSSL \cite{lazard2023giga}}  & 68.2 \scriptsize{$\pm$ 6.6} & 78.7 \scriptsize{$\pm$ 4.8} & 82.8 \scriptsize{$\pm$ 4.2} & 88.9 \scriptsize{$\pm$ 1.9} & -- & -- & -- & -- \\

& \multirow{1}{*}{GigaPath-Mean \cite{xu2024gigapath}}  & 59.7 \scriptsize{$\pm$ 6.6} & 69.8 \scriptsize{$\pm$ 5.2} & 77.0 \scriptsize{$\pm$ 6.1} & 85.8 \scriptsize{$\pm$ 3.4} & 64.8 \scriptsize{$\pm$ 9.7} & 79.7 \scriptsize{$\pm$ 5.3}  & 84.5 \scriptsize{$\pm$ 2.9}  & 91.7 \scriptsize{$\pm$ 1.9}  \\

& \multirow{1}{*}{GigaPath \cite{xu2024gigapath}}  & 58.7 \scriptsize{$\pm$ 6.7} & 68.6 \scriptsize{$\pm$ 4.9} & 75.4 \scriptsize{$\pm$ 4.8} & 84.0 \scriptsize{$\pm$ 2.5} & 64.0 \scriptsize{$\pm$ 11.5} & 78.0 \scriptsize{$\pm$ 7.5}  & 83.1 \scriptsize{$\pm$ 2.9}  & 90.3 \scriptsize{$\pm$ 2.0}  \\


& \multirow{1}{*}{$\ours$}  & 83.8 \scriptsize{$\pm$ 8.0} & 91.2 \scriptsize{$\pm$ 1.3} & 91.7 \scriptsize{$\pm$ 2.0} & 93.2 \scriptsize{$\pm$ 0.9} & 84.0 \scriptsize{$\pm$ 7.7} & 91.9 \scriptsize{$\pm$ 2.4} & 93.8 \scriptsize{$\pm$ 1.4} & \underline{96.0} \scriptsize{$\pm$ 0.8} \\

& \multirow{1}{*}{$\ours$-SE}  & \textbf{87.2} \scriptsize{$\pm$ 6.6} & \underline{93.2} \scriptsize{$\pm$ 1.0} & \underline{93.1} \scriptsize{$\pm$ 1.6} & 94.1 \scriptsize{$\pm$ 0.8} & \underline{85.6} \scriptsize{$\pm$ 7.2} & \textbf{93.7} \scriptsize{$\pm$ 1.7} & \textbf{95.3} \scriptsize{$\pm$ 0.9} & \textbf{96.7} \scriptsize{$\pm$ 0.4} \\

\midrule

\parbox[t]{3mm}{\multirow{7}{*}{\rotatebox[origin=c]{90}{\textbf{Prototyping}}}} 
& \multirow{1}{*}{\textsc{Mean}}  & 69.1 \scriptsize{$\pm$ 10.3} & 81.8 \scriptsize{$\pm$ 6.9} & 84.2 \scriptsize{$\pm$ 5.7} & 91.3 \scriptsize{$\pm$ 2.8} & 68.8 \scriptsize{$\pm$ 9.7} & 78.5 \scriptsize{$\pm$ 5.8} & 83.9 \scriptsize{$\pm$ 3.2} & 86.2 \scriptsize{$\pm$ 3.0} \\

& \multirow{1}{*}{\textsc{Intra}}  & 69.3 \scriptsize{$\pm$ 9.0} & 81.7 \scriptsize{$\pm$ 5.9} & 83.9 \scriptsize{$\pm$ 5.0} & 89.6 \scriptsize{$\pm$ 2.4} & 68.4 \scriptsize{$\pm$ 8.4} & 75.8 \scriptsize{$\pm$ 6.8} & 81.9 \scriptsize{$\pm$ 4.0} & 85.4 \scriptsize{$\pm$ 3.0} \\

& \multirow{1}{*}{HIPT$_{\text{CLS-4k}}$ \cite{chen2022scaling}}  & 62.1 \scriptsize{$\pm$ 4.1} & 68.3 \scriptsize{$\pm$ 6.2} & 73.6 \scriptsize{$\pm$ 6.9} & 78.7 \scriptsize{$\pm$ 2.4} & 66.3 \scriptsize{$\pm$ 12.3} & 75.7 \scriptsize{$\pm$ 7.3} & 79.1 \scriptsize{$\pm$ 4.2} & 82.3 \scriptsize{$\pm$ 0.9} \\

& \multirow{1}{*}{GigaSSL \cite{lazard2023giga}}  & 67.9 \scriptsize{$\pm$ 5.9} & 78.5 \scriptsize{$\pm$ 5.4} & 82.6 \scriptsize{$\pm$ 4.8} & 88.4 \scriptsize{$\pm$ 1.6} & -- & -- & -- & -- \\

& \multirow{1}{*}{GigaPath-Mean \cite{xu2024gigapath}}  & 58.8 \scriptsize{$\pm$ 6.2} & 70.0 \scriptsize{$\pm$ 5.7} & 73.5 \scriptsize{$\pm$ 7.0} & 81.7 \scriptsize{$\pm$ 4.1} & 65.7 \scriptsize{$\pm$ 9.5} & 78.8 \scriptsize{$\pm$ 4.9} & 81.6  \scriptsize{$\pm$ 2.7} & 85.5 \scriptsize{$\pm$ 2.0} \\

& \multirow{1}{*}{GigaPath \cite{xu2024gigapath}}  & 58.1 \scriptsize{$\pm$ 6.3} & 68.4 \scriptsize{$\pm$ 5.1} & 71.8 \scriptsize{$\pm$ 7.0} & 80.1 \scriptsize{$\pm$ 3.2} & 64.4 \scriptsize{$\pm$ 11.3} & 76.8 \scriptsize{$\pm$ 8.3} & 80.4  \scriptsize{$\pm$ 2.1} & 83.3 \scriptsize{$\pm$ 1.8} \\


& \multirow{1}{*}{$\ours$}  & 83.2 \scriptsize{$\pm$ 7.8} & 91.6 \scriptsize{$\pm$ 1.6} & 92.7 \scriptsize{$\pm$ 1.3} & \underline{94.4} \scriptsize{$\pm$ 0.6} & 84.6 \scriptsize{$\pm$ 8.4} & 91.1 \scriptsize{$\pm$ 3.0} & 93.1 \scriptsize{$\pm$ 1.4} & 94.9 \scriptsize{$\pm$ 0.9} \\

& \multirow{1}{*}{$\ours$-SE}  & \underline{86.4} \scriptsize{$\pm$ 6.6} & \textbf{93.4} \scriptsize{$\pm$ 1.0} & \textbf{94.0} \scriptsize{$\pm$ 1.1} & \textbf{95.0} \scriptsize{$\pm$ 0.4} & \textbf{86.3} \scriptsize{$\pm$ 7.7} & \underline{93.2} \scriptsize{$\pm$ 2.2} & \underline{94.9} \scriptsize{$\pm$ 0.8} & 95.8 \scriptsize{$\pm$ 0.8} \\

\bottomrule
\end{tabular}
}
\end{table*}

We showcase the performance of $\ours$ and $\oursSE$ (i.e., with stain encoding) on few-shot classification (Sec.~\ref{sec:fewshot}), and full classification (Sec.~\ref{sec:fullcls}), that we complement by a series of ablations (Sec.~\ref{sec:ablation}). We benchmark $\ours$ against four MIL methods: single head ABMIL~\cite{ilse2018attention}, TransMIL~\cite{shao2021transmil}, IB-MIL~\cite{li2023task} and ILRA~\cite{xiang2023exploring}; four intra-modal SSL methods: \textsc{Intra}, HIPT~\cite{chen2022scaling}, GigaSSL~\cite{lazard2023giga}, and GigaPath~\cite{xu2024gigapath} (work concurrent to $\ours$); and mean pooling (\textsc{Mean} and GigaPath-\textsc{Mean} constructed using the average of GigaPath patch embeddings). 

\subsection{Few-shot results} \label{sec:fewshot}

Fig.~\ref{fig:fig2} highlights the few-shot classification of $\ours$ against baselines (for each task: best MIL, best intra-modal, and \textsc{Mean}). Detailed morphological subtyping performance is reported in Table~\ref{tab:brca_fewshot}, molecular subtyping in Supplementary Table 3,
and kidney transplant rejection in Supplementary Table 7.

\noindent\textbf{$\ours$ \emph{vs.} rest} $\ours$ outperforms all baselines in 13/13 tasks, in some cases by a significant margin, \emph{e.g.,} +10.1\% over \textsc{Intra} in TCGA Breast ($k$=10, prototyping classification), or +9.0\% over ABMIL in BWH Breast ($k$=5, linear probing). This performance is achieved using simple downstream models based on linear probing or prototyping classification, whereas MIL methods are trained from scratch for each task.

\noindent\textbf{\textsc{Mean} \emph{vs.} \textsc{Intra}} Despite its simplicity, the \textsc{Mean} baseline offers high performance, often surpassing \textsc{Intra}, HIPT and GigaSSL. This highlights (1) the importance of powerful domain-specific patch encoders and (2) the complexity of deriving an information-rich training signal using information from the slide itself, further motivating the exploration of multimodal pretraining for slide representation learning. 

\noindent\textbf{MIL comparisons} Despite recent advances in MIL, ABMIL remains a strong baseline in a few-shot setting. In some cases, ABMIL is outperformed by IB-MIL~\cite{li2023task}. TransMIL, which includes patch-to-patch context using self-attention approximation, performs poorly, which we hypothesize is due to overfitting.

\noindent\textbf{$\ours$ \emph{vs.} \textsc{Intra}} $\ours$ outperforms the $\textsc{Intra}$ baseline for all values of $k$ on all tasks. This highlights the importance of using clinically and biologically meaningful ``views'' provided by multimodal pretraining. 

\noindent\textbf{$\ours$ \emph{vs.} GigaPath \emph{vs.} GigaPath-\textsc{Mean}} $\ours$ outperforms GigaPath on all tasks in both linear probing and prototyping evaluation, in most cases by a significant margin, e.g., +17.7\% in TCGA subtyping with linear probing ($k$=10). Interestingly, GigaPath-(\textsc{Mean}) (obtained by taking the average of the raw patch embeddings) reaches better performance than GigaPath slide encoder which suggests that intra-SSL can degrade performance, even when scaling to large models and number of samples. While GigaPath is a pan-cancer model, it was trained on a comparable number of breast samples (around 4,500 WSIs), which underscores (i) the quality of multistain pretraining, and (ii) the complexity of building pan-cancer models. 

\noindent\textbf{Generalization to other stains} As $\ours$ is stain-agnostic, we can use it for encoding non-H\&E stains. Specifically, we perform fine-tuning of $\ours$ multi-head encoder (\textsc{FineTune}) for quantification of ER and PR slides (framed as a 3-class and 6-class tasks) on the MGH cohort (Fig.~\ref{fig:fig3}.a and Supplementary Table 2).
We compare it against $\ours$ architecture trained from scratch and $\textsc{Mean}$. All models trained using $k$=25 examples per class. Fine-tuning leads to consistently better performance than random weight initialization, with a +7.5\% gain on 3-class ER and +5.6\% gain on 3-class PR quantification (Fig.~\ref{fig:fig3}.a and Supplementary Table 4).

\subsection{Full classification} \label{sec:fullcls}

\begin{figure*}[t]
   \centering
   \includegraphics[width=1.0\linewidth]{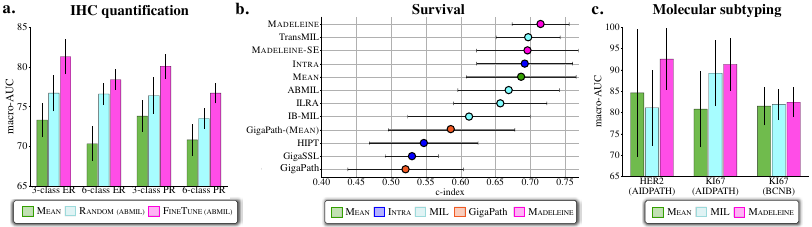}
   \caption{\textbf{Evaluation of $\ours$ and baselines on IHC quantification and survival prediction.}
   \textbf{a.} We fine-tune $\ours$ for IHC quantification on the MGH cohort (N=962 ER and N=1,071 PR slides). 3-class and 6-class variants are derived from IHC scores extracted in pathology reports. Models are trained with $k$=25 examples per class. \textsc{Random} uses $\ours$ architecture trained from scratch; \textsc{FineTune} is initialized with $\ours$ pretrained weights. We report the mean and standard deviation (std) on a 5-fold label-stratified train-test study. 
   \textbf{b.} Survival prediction on TCGA Breast (N=1,041 slides). We report mean and std using a 5-fold site-stratified cross-validation. ``SE'' is $\ours$ with stain encodings.
   \textbf{c.} Molecular subtyping of $\ours$ fine-tuned on AIDPATH (N=48 for HER2 and N=50 for KI67) and BCNB (N=1,058). Evaluation using 5-fold cross-validation. MIL refers to the best of four MIL baselines.
   }
   \label{fig:fig3}
\end{figure*}

\begin{table*}[t]
\centering
\caption{\textbf{Ablation study of $\ours$ loss.} Survival was evaluated using c-index and site-stratified 5-fold cross-validation. Subtyping and molecular status prediction were evaluated using macro-AUC and prototyping evaluation ($k$=25) repeated ten times with fixed seed across baselines. Standard deviation reported over the 10 runs. ``MSE'' stands for Mean-Squared Error.
Best in \textbf{bold}, second best is \underline{underlined}.}
\label{tab:ablation_loss}
\scalebox{1.0}{
\begin{tabular}{l|ccccc}
\toprule
Model/Data & \textbf{TCGA} & \textbf{BWH} & \textbf{TCGA} & \textbf{BCNB} & \textbf{Avg}\\
&\textbf{Survival} ($\uparrow$) & \textbf{Subtyping} ($\uparrow$) & \textbf{PR} ($\uparrow$) & \textbf{ER} ($\uparrow$) &\\
\midrule

\textsc{Mean} & 68.8 \scriptsize{$\pm$ 7.9} & 86.2 \scriptsize{$\pm$ 3.0} & 70.8 \scriptsize{$\pm$ 2.0} & 76.9 \scriptsize{$\pm$ 2.2} & 75.7 \\

\textsc{Intra} & 69.2 \scriptsize{$\pm$ 6.9} & 85.4 \scriptsize{$\pm$ 3.0} & 71.6 \scriptsize{$\pm$ 1.4} & 77.7 \scriptsize{$\pm$ 1.8} & 76.0 \\

\textsc{MSE} & 68.0 \scriptsize{$\pm$ 9.3} & 80.9 \scriptsize{$\pm$ 2.8} & 65.2 \scriptsize{$\pm$ 2.8} & 69.0 \scriptsize{$\pm$ 3.3} & 70.8 \\
 
\textsc{InfoNCE} & 69.9 \scriptsize{$\pm$ 8.1} & 93.3 \scriptsize{$\pm$ 0.9} & 74.5 \scriptsize{$\pm$ 1.3} & 82.8 \scriptsize{$\pm$ 1.7} & 80.1 \\
 
\textsc{GOT} & 70.1 \scriptsize{$\pm$ 3.6} & 85.9 \scriptsize{$\pm$ 2.6} & 70.1 \scriptsize{$\pm$ 2.0} & 75.8 \scriptsize{$\pm$ 2.7} & 75.5 \\
 
\textsc{InfoNCE} $+$ GOT & \textbf{71.5} \scriptsize{$\pm$ 4.1} & \textbf{94.9} \scriptsize{$\pm$ 0.9}  & \textbf{76.4} \scriptsize{$\pm$ 1.2} & \underline{83.0} \scriptsize{$\pm$ 1.6} & \textbf{81.5} \\

\textsc{InfoNCE} $+$ GOT $+$ \textsc{Intra} & \underline{71.0} \scriptsize{$\pm$ 6.2} & \textbf{94.9}  \scriptsize{$\pm$ 1.1} & \textbf{76.4} \scriptsize{$\pm$ 1.2}  & \textbf{83.3} \scriptsize{$\pm$ 1.3} & \underline{81.4} \\

\bottomrule
\end{tabular}
}
\end{table*}

\begin{figure*}[t]
   \centering
   \includegraphics[width=1.0\linewidth]{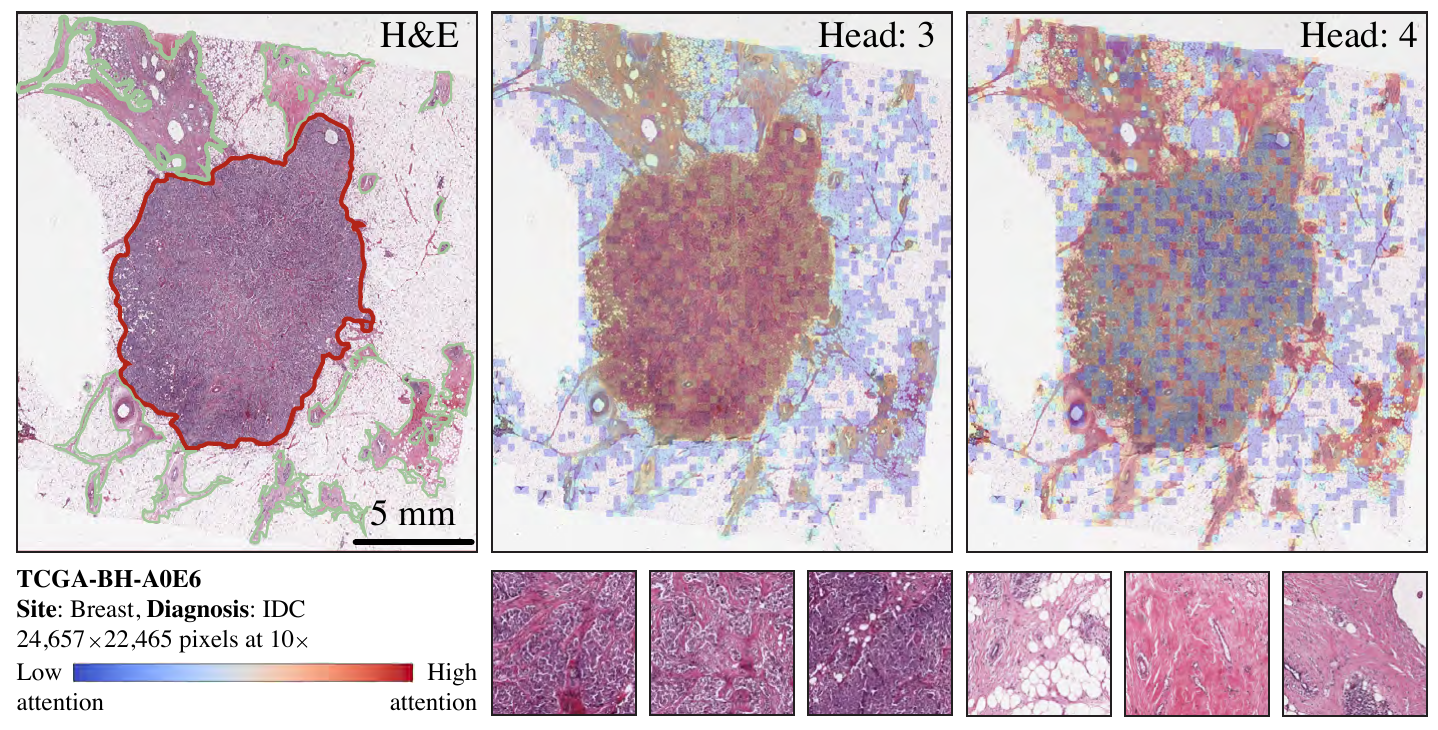}
   \caption{\textbf{$\ours$ attention weight visualization in a breast cancer case.}
   Attention weights of the third (focusing on tumor, annotated in red) and fourth (focusing on non-tumor regions, annotated in green) heads of $\ours$ slide encoder along with high attention patches per head.}
   \label{fig:fig4}
\end{figure*}

Beyond few-shot classification, we assess $\ours$ in a supervised setting using 5-fold cross-validation, where we directly use $\ours$ embeddings for survival prediction and molecular subtyping.

\noindent\textbf{Survival.} We perform survival outcome prediction on TCGA Breast using a 5-fold site-stratified cross-validation. $\ours$ and other slide-level models (HIPT, GigaSSL and \textsc{Intra}) are trained using a Cox proportional hazards objective from the slide embedding. All MIL models are trained with a survival NLL objective following prior work~\cite{chen2022pan, jaume2023modeling}. $\ours$ leads to the best survival prediction reaching 0.71 c-index outperforming all baselines (Fig.~\ref{fig:fig3}.b and Supplementary Table 5).

\noindent\textbf{Molecular subtyping.} In addition, we use logistic regression to predict HER2 status in AIDPATH and KI67 status in AIDPATH and BCNB from H\&E. In AIDPATH, $\oursSE$ leads to +11.4\% performance boost over the best MIL in HER2, and +1.8\% in KI67 (Fig.~\ref{fig:fig3}.c and Supplementary Table 6.

\subsection{Ablation} \label{sec:ablation}

All ablations were run using $\ours$ pretrained on breast cancer slides, evaluated using AUC, and benchmarked using prototyping classification ($k$=25) on a set of three representative tasks: (1) BWH Breast subtyping, (2) TCGA PR classification, and (3) BCNB ER classification. In addition, we benchmark TCGA Breast survival using a Cox model trained using 5-fold cross-validation. Prototyping is not sensitive to hyper-parameter selection compared to linear probing, making it ideal for ablating components of $\ours$.

\textbf{Loss ablation.} We perform a thorough ablation of $\ours$ loss function by retraining models with \textsc{InfoNCE} alone, cross-modal Mean-Squared Error (replacing \textsc{InfoNCE}), \textsc{GOT} alone, combining \textsc{InfoNCE} and \textsc{GOT} ($\ours$ default), and finally combining \textsc{InfoNCE}, \textsc{GOT} and \textsc{Intra} (Table~\ref{tab:ablation_loss}).
\textsc{InfoNCE} alone significantly outperforms \textsc{Mean} (+4.4\% AUC), \textsc{Intra} (+4.1\%) and MSE (+9.3\%). \textsc{GOT} alone performs similarly to \textsc{Mean} and \textsc{Intra}. When combining \textsc{InfoNCE} and \textsc{GOT}, we observe an additional gain of +1.4\% over \textsc{InfoNCE}. However, including an \textsc{Intra} objective on top leads to a similar performance. Overall, the global cross-modal \textsc{InfoNCE} objective remains the most critical component, which benefits from the local GOT cross-modal objective. 

\textbf{Feature extractor ablation.} We further test if the benefits of $\ours$ pretraining generalize when using CTransPath\cite{wang2021transpath}, a state-of-the-art patch encoder based on the Swin-Transformer model and that was pretrained on 15 million patches from TCGA and PAIP (Supplementary Table 8).
When comparing the \textsc{Mean} baseline, our patch encoder significantly outperforms CTransPath (+5.0\% AUC). The same observation holds using $\ours$ embeddings, where using our patch encoder leads to +10.5\% AUC gain over CTransPath. Overall, these results further assert that (1) using powerful domain-specific feature encoders trained on large amounts of diverse data is necessary, and (2) $\ours$ pretraining leads to high performance even when using weaker feature encoders. 

\textbf{Architecture ablation.} $\ours$ explores two architectural features: (1) the use of a multi-head (MH) attention network and (2) the use of a learnable stain encoding (Supplementary Table 9).
Using multiple ABMIL heads (four in $\ours$) leads to a consistent performance gain (on average of +2.1\%). We hypothesize the gain arises from each head focusing on different morphologies during pretraining. Adding stain-encoding does not have a consistent effect, as it boosts performance in morphological subtyping but decreases performance in molecular subtyping. Replacing the ABMIL architecture with a TransMIL backbone~\cite{shao2021transmil} leads to lower performance (on average -2.9\% over ABMIL and -5.0\% over MH-ABMIL).

\subsection{$\ours$ attention visualization}

By visualizing head-specific attention weights, we can gain insights into the internal behavior of $\ours$ (Fig.~\ref{fig:fig4}). We show that different heads learn to focus on morphologically distinct regions, \emph{e.g.,} Head-3 focuses on tumor while Head-4 focuses on non-tumor stroma. This is a remarkable finding as $\ours$ was not given any morphological labels like tumor grade or subtype during training. Additional example heatmaps are provided in Supplementary 7.

\section{Conclusion}

In this study, we present $\ours$, a method exploring multimodal pretraining for \emph{slide representation learning} based on multistain alignment. Our method utilizes extensive datasets of multistain slides, where we consider each stain as a unique perspective of the standard H\&E-stained slide, each revealing different aspects of the tissue's biological state. We demonstrate that $\ours$ slide encoder outperforms multiple instance learning and intra-modal pretraining models in few-shot and full classification scenarios across various tasks, ranging from morphological and molecular subtyping to prognosis prediction to IHC quantification. Our method currently incorporates four to five different stains per sample, yet in clinical practice, more stains can be available to assist pathologists. This opens up promising avenues for expanding $\ours$ pretraining to include a broader range of stains. Furthermore, while our focus has been on multimodal pretraining with multiple stains, there exists a potential to explore other spatial modalities, such as those based on immunofluorescence, mass spectrometry, or spatial transcriptomics~\cite{jaume2024hest} for slide representation learning.

\clearpage
\bibliographystyle{splncs04}
\bibliography{main}

\setcounter{section}{0}
\setcounter{table}{0}
\setcounter{figure}{0}
\setcounter{equation}{0}

\captionsetup[figure]{name={Supp. Fig.}}
\captionsetup[table]{name={Supp. Table}}
\renewcommand{\theequation}{Supp. Eq. \arabic{equation}}


\clearpage
\begin{center}
    \LARGE\textbf{Supplementary Material}\\[1em]
    \large\textbf{Multistain Pretraining for Slide Representation \\ Learning in Pathology}\\[0.5em]
\end{center}





\noindent We provide complementary information on the model architecture and training, additional results, and interpretability examples: 

\begin{enumerate}
    \item \textbf{Sec. \ref{supp:imp_deets}}: Implementation details around pretraining, aggregator architecture, and graph optimal transport loss.

    \item \textbf{Sec. \ref{supp:datasets}}: Detailed descriptions of datasets used for pretraining and the downstream evaluations.

    \item \textbf{Sec. \ref{supp:baselines}}: Supervised Multiple Instance Learning (MIL) baselines.

    \item \textbf{Sec. \ref{supp:add_breast}} and \textbf{\ref{supp:add_kidney}}: Additional results on downstream breast and kidney tasks.

    \item \textbf{Sec. \ref{supp:add_ablations}}: Ablations of loss and aggregator architecture.

    \item \textbf{Sec. \ref{supp:add_interpret}}: Additional interpretability examples of breast cancer cases. 

    \item \textbf{Sec. \ref{supp:limitations}}: Limitations of $\ours$.
\end{enumerate}

\section{Implementation details}
\label{supp:imp_deets}

\subsection{Contrastive loss}

Formally, we define a batch of $B$ cases, where each case includes $K$ pairs $(\textbf{h}^{\he}_i, \textbf{h}^{s_k}_i)_{k=1}^{K}$, where $s_k$ represents a non-H\&E stain. The objective $\mathcal{L}_{\textsc{infoNCE}}$ is given by: 
\begin{equation}
\label{eq:loss_symcl}
\begin{split}
\mathcal{L}_{\textsc{infoNCE}} = -\frac{1}{K} \sum_{k=1}^{K}\frac{1}{2B} \Biggl( &\sum_{b=1}^{B} \log \frac{\exp \left(\frac{1}{\tau} \big(\boldsymbol{h}_b^{\he}\big)^{\text{T}} \boldsymbol{h}^{s_k}_{b}\right)}{\sum_{b'=1}^{B} \exp \left(\frac{1}{\tau}  \big(\boldsymbol{h}_b^{\he}\big)^\text{T}  \boldsymbol{h}^{s_k}_{b'}\right)} \\
& + \sum_{b=1}^{B} \log \frac{\exp \left(\frac{1}{\tau} \big(\boldsymbol{h}^{s_k}_b\big)^{\text{T}}  \boldsymbol{h}_{b}^{\he}\right)}{\sum_{b'=1}^{B} \exp \left(\frac{1}{\tau}  \big(\boldsymbol{h}^{s_k}_b\big)^{\text{T}}  \boldsymbol{h}_{b'}^{\he}\right)} \Biggl), \\
\end{split}
\end{equation}
where the first and second terms represent the H\&E-to-$s_k$ and $s_k$-to-H\&E contrastive loss, respectively. $\tau$ represents the Softmax temperature parameter.

\subsection{Multi-head attention architecture}
\label{app:multihead}
$\ours$ uses a multi-head attention-based Multiple Instance Learning (MIL) architecture. Before applying each head, the patch embeddings are passed through a common pre-attention network consisting of 3 layers with 512 hidden units, layer normalization, GELU activation, and 0.1 dropout. Each attention head comprises a gated-attention network, consisting of a 2-layer MLP with 512 hidden units with Softmax activation and 0.25 dropout. The attention score $a_{i,j}^{m}$ for each patch derived from the $m^{th}$ attention head of $M$ total heads are defined as:

\begin{equation}
\label{eq:mha}
    a_{i,j}^m = \frac{\exp\big(\mathbf{w}_m( \tanh(\mathbf{V}_m\widetilde{\mathbf{H}}_{i,j}^{\text{HE}})\odot\operatorname{sigm}(\mathbf{U}_m\widetilde{\mathbf{H}}_{i,j}^{\text{HE}})\big)}{\sum_{j'=1}^{N_\he}\exp\big(\mathbf{w}_m( \tanh(\mathbf{V}_m\widetilde{\mathbf{H}}_{i,j'}^{\text{HE}})\odot\operatorname{sigm}(\mathbf{U}_m\widetilde{\mathbf{H}}_{i,j'}^{\text{HE}})\big)},\,\,\,\forall m.
\end{equation}

The output of each head is concatenated, and a post-attention network consisting of two linear layers with 2048 and 512 units is applied to get a slide embedding for each stain.

\subsection{Additional information on the GOT objective} \label{app:loss}

Additional details of the Graph Optimal Transport objectives are as follows,

\paragraph{1) Graph building:} Each stain-specific graph is defined by instantiating 256 randomly sampled patches as nodes from the slide (sampling is done as each slide can have $>10,000$ patches, making it computationally infeasible to calculate the complete optimal transport-based loss). Then, an edge is built between two nodes (\emph{i.e.,} two patches) if the cosine similarity between their patch embeddings is larger than a threshold. The threshold is dynamically constructed and is set at the lowest similarity value, increasing by 0.1 times the difference between the highest and lowest similarity values.

\paragraph{2) $\mathcal{L}_{\text{WD}}$: } Denoting $\mathbf{T}\in\mathbb{R}_{+}^{N_{\he}\times N_{s_k}}$ as the transport plan, we can minimize the Wasserstein Distance (WD) between distributions $\hat{p}_{\he}$ and $\hat{p}_{s_k}$ by finding the optimal transport plan
\begin{equation}
    \mathcal{L}_{\text{WD}}(\hat{p}_{\text{HE}}, \hat{p}_{s_k}) = \min_{\textbf{T}}\sum_{j}\sum_{m} \mathbf{T}_{j,m}\cdot C(v^{\he}_j, v^{s_k}_m),
\end{equation}
such that $\sum_{j=1}^{N_{\he}}\mathbf{T}_{j,m}=1/N_{s_k}, \forall m$ and $\sum_{m=1}^{N_{s_k}}\mathbf{T}_{j,m}=1/N_{\he}, \forall j$. The cost between cross-modal embeddings $C(v^{\he}_j, v^{s_k}_m)$ is computed with the cosine distance metric. 

\paragraph{3) $\mathcal{L}_{\text{GWD}}$: } In addition to the node-matching with $\mathcal{L}_{\text{WD}}$, we also wish to match the graph topology via comparing the edge distance between stain-specific graphs. Denoting $\widetilde{\mathbf{T}}\in\mathbb{R}_{+}^{N_{\he}\times N_{s_k}}$ as the transport plan as before,
\begin{equation}
    \mathcal{L}_{\text{GWD}}(\hat{p}_{\text{HE}}, \hat{p}_{s_k}) = \min_{\textbf{T}}\sum_{j,j',m,m'} \widetilde{\mathbf{T}}_{j,m}\widetilde{\mathbf{T}}_{j',m'}\cdot C(v^{\he}_j, v^{s_k}_m, v^{\he}_{j'}, v^{s_k}_{m'}),
\end{equation}
such that $\sum_{j=1}^{N_{\he}}\widetilde{\mathbf{T}}_{j,m}=1/N_{s_k}, \forall m$ and $\sum_{m=1}^{N_{s_k}}\widetilde{\mathbf{T}}_{j,m}=1/N_{\he}, \forall j$. The cost between the pairs $(v^{\he}_j, v^{\he}_{j'})$ and $(v^{s_k}_m, v^{s_k}_{m'})$ is given as $C(v^{\he}_j, v^{s_k}_m, v^{\he}_{j'}, v^{s_k}_{m'})=\lVert c(v^{\he}_j, v^{\he}_{j'}) - c(v^{\he}_m, v^{\he}_{m'})\rVert$, with $c(\cdot,\cdot)$ representing the cosine similarity metric.


\subsection{$\ours$ pretraining }
\label{supp:pretraining}
We pretrained $\ours$ with AdamW optimizer and a batch size of 90 for 120 epochs. The learning rate is linearly ramped up during a 5-epoch warmup from 1e-9 to 1e-4. Then, we employed a cosine scheduler to reach the final learning rate of 1e-8 after 120 epochs. To increase training diversity and simplify batch processing, we sample a fixed and random subset of patches per slide, specifically 2048 patch embeddings. In slides with fewer patches, we perform random over-sampling. All training settings are summarized in Appendix Table \ref{tab:hparams}.

\begin{table*}
  \centering
  \caption{
  \textbf{$\ours$ pretraining and architectural hyperparameters.} 3 $\times$ 24GB NVIDIA 3090Ti GPUs were used for training. Batch size refers to the total batch size across all GPUs.}
  \begin{tabular}{l|l}
    \toprule
    Hyperparameter & Value \\
    \midrule
    Heads & 4 \\
    Head activation & GELU \\
    Patch embedding dimension & 512 \\
    Pre-attention hidden dimension & 512 \\
    Patches sampled (training) & 2048 \\
    Stain encoding dimension & 32 \\
    \midrule
    AdamW $\beta$ & (0.9, 0.999) \\
    Batch size & 90 \\
    Warmup epochs & 5 \\
    Max epochs & 120 \\
    Learning rate schedule & Cosine \\
    Learning rate (start) & 0 \\
    Learning rate (post warmup) & 1e-4\\
    Learning rate (final) & 1e-8 \\
    Weight decay & 0.01 \\
    \textsc{infoNCE} Temperature & 0.001 \\
    Patches sampled for GOT & 256 \\
    GOT $\gamma$ & 0 \\
    Automatic mixed precision & bloaft16 \\
    Early stopping criteria & SmoothRank \cite{garrido2022rankme} \\
    \bottomrule
  \end{tabular}
  \label{tab:hparams}
\end{table*}

\clearpage
\subsection{Early stopping with rank analysis}
\label{supp:rank}
Following~\cite{garrido2022rankme}, we use the rank as a measure of the quality of the underlying latent space learned during $\ours$ pretraining and save the model weights from the highest rank iteration (no models are saved during the first 20 epochs of training). We compute the rank as the entropy of the $d$ (assuming $d<n$) L1-normalized singular values of the slide embedding matrix $H \in \real^{n \times d}$. Specifically, we have:
\begin{align}
    \text{RankMe}(H) &= \exp(-\sum_{k=1}^d p_k\log(p_k))\;, \\
    p_k &= \frac{\sigma_k(H)}{|\sigma(H)|_1} + \epsilon
\end{align}
where $\sigma_k$ denotes the $k-$th singular of $H$ (sorted from large to low), and $\epsilon$ is small constant set to $1e-7$ for numerical stability.

\subsection{Additional information on evaluation} \label{app:eval}

Few-shot evaluation is based on linear probing and prototyping classification. 

\textbf{Linear probing} Linear probing is implemented using a logistic regression objective based on sklearn. We use the default sklearn L2 regularization (set to 1.0) with an lbfgs solver. We set the maximum number of training iterations to 10,000 for all experiments.  

\textbf{Prototyping} We define positive and negative slide ``prototypes'' $p^{+}, p^{-}$ as the average of $k$ ($k$=1,5,10,25) slide embeddings using downstream task labels. Subsequently, we measure the similarity between a query slide embedding $q_i$ and the two prototypes using the L2 distance. We apply this evaluation for morphological and molecular subtyping, and allograft rejection prediction.

\clearpage
\section{Datasets} 
\label{supp:datasets}

Overall, our study comprises a total of 23,580 whole slide images (WSIs) from two organs (breast and kidney) and includes eight different immunohistochemistry and special stains. We use 16,281 of these WSIs for pretraining and 7,299 for downstream evaluation. We now detail all the sources of the WSIs used in the study. \\

\noindent \textbf{Acrobat (multi-stain, pretraining)} Acrobat is a multi-stain dataset originally proposed as part of the AutomatiC Registration Of Breast cAncer Tissue MICCAI challenge \cite{weitz2022acrobat,weitz2023multi}. It comprises 4,211 whole slide images (WSI) sourced from 1,153 patients diagnosed with primary breast cancer. These WSIs are available at a magnification of $10\times$ (equivalent to 1$\mu$m/px) and show tissue resections, which have been processed using either hematoxylin and eosin (H\&E) staining or immunohistochemistry (IHC). For every patient included in the dataset, there is one WSI that has been stained with H\&E, along with a minimum of one and a maximum of four WSIs of tissue from the same tumor that has been stained with ER (N=844), PR (N=837), HER2 (N=534), or KI67 (N=843). The collection of slides was digitized at Karolinska Institutet in Stockholm, Sweden, during routine clinical workflows. Data can be downloaded at \url{https://acrobat.grand-challenge.org/data/}.\\

\noindent \textbf{TCGA Breast (H\&E, downstream)} We collected N=1,041 primary cases from the TCGA Breast Invasive Carcinoma (BRCA) cohort, which comprises N=831 Invasive Ductal Carcinoma (IDC) and N=210 Invasive Lobular Carcinoma (ILC). For each case, we downloaded the corresponding disease-specific survival and associated censorship status, subtype (IDC and ILC), and molecular status: ER (N=996; 780 positive, 216 negative), PR (N=993; 678 positive, 315 negative), and HER2 (N=693; 158 positive, 535 negative) from UCSC Xena~\cite{goldman2020visualizing} and cBioPortal~\cite{cerami2012cbio}. WSIs can be downloaded from Genomics Data Commons ({\url{https://portal.gdc.cancer.gov/}}).\\

\noindent \textbf{BCNB (H\&E, downstream)} The Cancer Core-Needle Biopsy WSI (BCNB) dataset comprises N=1,058 patients, with a single side associated with each patient \cite{xu2021predicting}. BCNB includes the molecular status of each patient: ER (N=1,058; 831 positive, 227 negative), PR (N=1,058; 790 positive, 268 negative), HER2 (1,058; 277 positive, 781 negative), and KI67 (1,058; 156 positive, 902 negative). The dataset was originally collected from hospital systems in Beijing, China, and is made publicly available at (\url{https://paperswithcode.com/dataset/bcdalnmp}). \\

\noindent \textbf{AIDPATH (H\&E, downstream)} AIDPATH dataset contains 50 breast cancer WSIs stained with H\&E. The dataset additionally provides HER2 expression (positive or negative, where equivocal cases are analyzed with FISH) (7 positive, 41 negative) and KI67 expression (provided as a percentage). We convert the continuous KI67 expression values to a binary task using $50\%$ as a threshold for IHC status prediction (19 positive, 31 negative). The dataset is made publicly available at \url{https://mitel.dimi.uniud.it/aidpath-db}. \\

\noindent \textbf{BWH Breast (H\&E, downstream)} We collected an invasive breast cancer cohort (N=1,265) from the archives of Hopsital-A, which comprises N=982 IDC and N=283 ILC cases. All cases were primary breast cancers and included resections and biopsies. All slides were scanned at $20\times$ or $40\times$ magnification. Using patient reports, we additionally collected molecular status: ER (N=874; 613 positive, 261 negative), PR (N=874; 504 positive, 370 negative), and HER2 (N=816; 151 positive, 665 negative). \\

\noindent\textbf{MGH Breast (Estrogen and Progesterone receptor stains, downstream)} We use another private breast cohort for IHC quantification of ER abundance (N=962) and PR abundance (N=1,071). We frame both ER and PR quantification as 3- and 6-class problems. For a detailed breakdown, see Appendix Table \ref{tab:hospB}. \\

\noindent\textbf{BWH Kidney (multi-stain, pretraining)} We collected a private renal transplant cohort comprising kidney biopsies from 1,069 renal transplant cases. Each case includes one to three tissue blocks, where each block consists of one to two H\&E-stained ($N$=4,638) and one to two periodic acid-Schiff (PAS) (N=4,630) slides, one Jones-stained slide (N=2,326) and one Trichrome-stained slide (N=2,328). In total, each case includes 6 to 18 slides. We held out 20\% of the cohort (210 cases, 463 H\&E slides) as an independent test set and used the rest for $\ours$ pretraining. All slides were processed at 20$\times$. We use H\&E slides of the held-out cases to screen for Antibody-mediated rejection (AMR, 2 class; 107 positive, 356 negative) and quantify Interstitial Fibrosis and Tubular Atrophy (IFTA, 3 classes; $\text{mild}: 292, \text{moderate}: 104, \text{advanced}: 67$). As each case includes several H\&E slides, we define two sub-tasks: ``single-slide" prediction, where we use a single slide per case (N=463 H\&E slides), and ``all-slides" prediction, where we use all available slides per case (N=305 H\&E slides).

\begin{table*}
\centering
\caption{\textbf{MGH Breast label distribution.}}
\label{tab:hospB}
\scalebox{1.0}{
\begin{tabular}{l|cc|cc}
\toprule
Model/Data & \multicolumn{2}{c}{\textbf{ER}}  & \multicolumn{2}{c}{\textbf{PR}} \\
& $\mathcal{C}=6$ & $\mathcal{C}=3$ & $\mathcal{C}=6$ & $\mathcal{C}=3$ \\
\midrule
$0$ & 175 & \multirow{2}{*}{335} & 168 & \multirow{2}{*}{337} \\
$<1\%$ & 160 &  & 169 &  \\
$1-10\%$ & 120 & \multirow{2}{*}{292} & 219 & \multirow{2}{*}{389} \\
$10-50\%$ & 172 &  & 170 & \\
$50-90\%$ & 176 & \multirow{2}{*}{335} & 175 & \multirow{2}{*}{344} \\
$>90\%$ & 159 &  & 169 & \\
\bottomrule
\end{tabular}
}
\end{table*}

\section{Baselines} \label{supp:baselines}
\subsection{Supervised multiple instance learning baselines}

We provide a detailed description of the four multiple-instance learning (MIL) approaches used in the study. 

\begin{enumerate}
    \item \textbf{ABMIL} \cite{ilse2018attention}: Attention-based multiple instance learning (ABMIL) is a popular MIL architecture. ABMIL operates as follows: first, it assigns patch-level importance scores through a gated-attention mechanism. Attention scores are used to weigh patch embeddings, which are subsequently summed to build a slide representation used for classification.
    
    \item \textbf{TransMIL} \cite{shao2021transmil}: Transformer-based multiple instance learning (TransMIL) replaces the gated attention from ABMIL with a low-rank Transformer. TransMIL first squares the sequence of low dimensional representations then applies a Pyramidal Positional Encoding module to encode spatial information and finally uses Nystrom attention \cite{xiong2021nystrom} to approximate self-attention scores between patches. The CLS token is finally taken as the slide-level representation.

    \item \textbf{Information Bottleneck MIL} \cite{li2023task}: Information bottlenecks (IB) are used to compress the WSI by removing uninformative instances (patch embeddings). IB aims to find patch instances that minimize the mutual information between the distribution of patches and patch representations. By only keeping such instances, \cite{li2023task} postulate that the most informative patches can be retained, which can then be aggregated into a slide embedding. 
    
    \item \textbf{Low-rank MIL} \cite{xiang2022exploring}: While TransMIL tries to learn slide-level representations by encoding patch correlations, it does not leverage the redundancy in WSIs, which \cite{xiang2022exploring} used to propose iterative low-rank attention (ILRA). Each ILRA block consists of two layers: one aims to project the sequence of patch representations to a low-rank space by cross-attending it with a latent matrix, and the second reconstructs the input. Performing max-pooling over the output of $k$ such layers yields a low-rank slide-level representation.  

\end{enumerate}

\subsection{Slide-level baselines}

\begin{enumerate}
    \item \textbf{\textsc{Mean}}: The \textsc{Mean} baseline is defined by taking the arithmetic average of the patch embedding constituting the slides.
    

    \item \textbf{HIPT} \cite{chen2022scaling}: Hierarchical Image Pyramid Transformer (HIPT) proposes a 3-level slide encoding schema, where each level is independently trained with a Transformer. The first level transforms patches into patch embeddings, which are then aggregated in region embeddings and finally into a slide embedding. 
    
    \item \textbf{GigaSSL} \cite{lazard2023giga}: GigaSSL, similar to the \textsc{Intra} baseline, is a method for learning slide representations based on different views of the same slide. It creates different views of a slide by sampling patches and applying augmentations such as random cropping. The different views are then pulled using a contrastive loss to learn the slide representation. Author-provided GigaSSL slide embeddings for TCGA Breast were taken from \url{https://data.mendeley.com/datasets/d573xfd9fg/3}.   

    \item \textbf{GigaPath} \cite{xu2024gigapath}: GigaPath is a concurrent work to ours scaling intra-SSL to large cohorts. It includes its own pan-cancer patch encoder that was pretrained on 171,000+ WSIs ($>$ 30,000 patients) using DINOv2. The slide encoder was trained using a LongNet model with masked auto encoding. We used the official GigaPath demo\footnote{https://github.com/prov-gigapath/prov-gigapath/blob/main/demo/run$\_$gigapath.ipynb} using (1) 256$\times$256-pixel patching at 20$\times$ and the latest HuggingFace tile and slide encoders. GigaPath-(\textsc{Mean}) is defined by taking the average of all patch embeddings, and GigaPath is defined by building a slide embedding using global pooling of all LongNet tokens at the 11th Transformer layer.  
\end{enumerate}

\clearpage
\section{Additional Breast results}
\label{supp:add_breast}

\begin{table*}
\centering
\caption{\textbf{Few-shot molecular status prediction from H\&E in TCGA Breast.} Evaluation using Macro-AUC. Standard deviation reported over ten runs. All results for $k=25$ training samples per class. $\ours$ refers to \textsc{InfoNCE + GOT}. Besides HIPT and GigaSSL, all models use the \emph{same} patch encoder. GigaSSL embeddings for BWH and BCNB cohorts were not available. Best in \textbf{bold}, second best is \underline{underlined}.}
\label{tab:brca_molstatus}
\scalebox{0.95}{
\begin{tabular}{ll|ccc|ccc|ccc}
\toprule
& Model/Data & \multicolumn{3}{c}{\textbf{TCGA} ($\uparrow$)}  & \multicolumn{3}{c}{\textbf{BCNB} ($\uparrow$)} & \multicolumn{3}{c}{\textbf{BWH} ($\uparrow$)}\\
& & ER &PR & HER2 & ER & PR & HER2 & ER & PR & HER2 \\
\midrule
\parbox[t]{3mm}{\multirow{8}{*}{\rotatebox[origin=c]{90}{\textbf{MIL}}}} 
& \multirow{1}{*}{ABMIL \cite{ilse2018attention}}  & 82.7 & 72.8  & 62.4  & 81.4  & 75.9 & 64.5 & 65.4 & 62.6 & 57.0 \\
& & \scriptsize{$\pm$ 3.1} & \scriptsize{${\pm}$ 2.5}  & \scriptsize{${\pm}$ 3.3}  & \scriptsize{${\pm}$ 4.5}  & \scriptsize{${\pm}$ 2.9} & \scriptsize{${\pm}$ 3.4}  & \scriptsize{${\pm}$ 3.5} & \scriptsize{${\pm}$ 2.6} & \scriptsize{${\pm}$ 2.4} \\

& \multirow{1}{*}{TransMIL \cite{shao2021transmil}}  & 75.1 & 63.5  & 55.0  & 68.7  & 63.4 & 56.0 & 54.8 & 54.3  & 51.8 \\
& & \scriptsize{${\pm}$ 5.5} & \scriptsize{${\pm}$ 7.1} & \scriptsize{${\pm}$ 2.9}  & \scriptsize{${\pm}$ 7.0}  & \scriptsize{${\pm}$ 8.2} & \scriptsize{${\pm}$ 4.1}  & \scriptsize{${\pm}$ 4.2} & \scriptsize{${\pm}$ 3.0} & \scriptsize{${\pm}$ 2.7} \\

& \multirow{1}{*}{IB-MIL \cite{li2023task}}  & 81.6 & 73.1  & \underline{62.6} & 81.0  & 76.7 & 64.3 & 65.8 & 61.7  & 55.2 \\
& & \scriptsize{${\pm}$ 2.8} & \scriptsize{${\pm}$ 2.5}  & \scriptsize{${\pm}$ 2.6}  & \scriptsize{${\pm}$ 2.8}  & \scriptsize{${\pm}$ 2.2} & \scriptsize{${\pm}$ 4.7}  & \scriptsize{${\pm}$ 3.1} & \scriptsize{${\pm}$ 2.5} & \scriptsize{${\pm}$ 3.7} \\

& \multirow{1}{*}{ILRA \cite{xiang2022exploring}}  & 82.3 & 74.0  & 62.2  & 79.0  & 74.2 & 63.6 & 63.2 & 59.9  & 55.1 \\
& & \scriptsize{${\pm}$ 2.7} & \scriptsize{${\pm}$ 2.4}  & \scriptsize{${\pm}$ 2.6}  & \scriptsize{${\pm}$ 7.2}  & \scriptsize{${\pm}$ 1.9} & \scriptsize{${\pm}$ 3.9}  & \scriptsize{${\pm}$ 3.9} & \scriptsize{${\pm}$ 4.7} & \scriptsize{${\pm}$ 3.5} \\

\midrule

\parbox[t]{3mm}{\multirow{16}{*}{\rotatebox[origin=c]{90}{\textbf{Linear probe}}}} & \multirow{1}{*}{\textsc{Mean}}  & 79.4 & 68.5 & 59.6 & 78.5 & 72.4 & 64.7 & 62.8 & 60.7 & 55.8 \\
& & \scriptsize{${\pm}$ 4.7} & \scriptsize{${\pm}$ 4.2} & \scriptsize{${\pm}$ 3.1} & \scriptsize{${\pm}$ 3.2} & \scriptsize{${\pm}$ 3.8} & \scriptsize{${\pm}$ 4.3} & \scriptsize{${\pm}$ 2.8} & \scriptsize{${\pm}$ 3.6} & \scriptsize{${\pm}$ 4.3} \\

& \multirow{1}{*}{Intra}  & 80.1 & 70.2 & 60.0 & 79.6 & 75.1 & 67.1 & 64.0 & 60.9 & 55.9 \\
& & \scriptsize{${\pm}$ 3.3} & \scriptsize{${\pm}$ 3.2} & \scriptsize{${\pm}$ 2.5} & \scriptsize{${\pm}$ 2.7} & \scriptsize{${\pm}$ 2.7} & \scriptsize{${\pm}$ 3.7} & \scriptsize{${\pm}$ 2.2} & \scriptsize{${\pm}$ 3.1} & \scriptsize{${\pm}$ 3.4} \\

& \multirow{1}{*}{HIPT$_{\text{CLS-4k}}$ \cite{chen2022scaling}}  & 74.1 & 63.9 & 61.9 & 65.3 & 62.2 & 54.0 & 60.3 & 57.0 & 52.9 \\
& & \scriptsize{${\pm}$ 3.0} & \scriptsize{${\pm}$ 2.9} & \scriptsize{${\pm}$ 3.7} & 3.7 & 4.4 & 3.0 & \scriptsize{${\pm}$ 3.6} & \scriptsize{${\pm}$ 3.6} & \scriptsize{${\pm}$ 3.7} \\

& \multirow{1}{*}{GigaSSL \cite{lazard2023giga}}  & 77.7 & 69.4 & 59.9 & -- & -- & -- & -- & -- & -- \\
& & \scriptsize{${\pm}$ 3.1} & \scriptsize{${\pm}$ 2.8} & \scriptsize{${\pm}$ 3.2} & -- & -- & -- & -- & -- & -- \\

& \multirow{1}{*}{GigaPath-(\textsc{Mean}) \cite{xu2024gigapath}}  & 77.7 & 68.1 & 58.3 & 76.2 & 71.0  & 63.7 & 64.6 & 59.3 & 54.4 \\
& & \scriptsize{${\pm}$ 4.8} & \scriptsize{${\pm}$ 2.8}  & \scriptsize{${\pm}$ 3.4} & \scriptsize{${\pm}$ 3.9} & \scriptsize{${\pm}$ 3.9} & \scriptsize{${\pm}$ 3.8} & \scriptsize{${\pm}$ 2.2} & \scriptsize{${\pm}$ 3.1} & \scriptsize{${\pm}$ 3.4} \\

& \multirow{1}{*}{GigaPath \cite{xu2024gigapath}}  & 76.0 & 66.7 & 57.4 & 74.1 & 68.9  & 61.8 & 63.0 & 58.2 & 53.2 \\
& & \scriptsize{${\pm}$ 5.0} & \scriptsize{${\pm}$ 2.7}  & \scriptsize{${\pm}$ 3.4} & \scriptsize{${\pm}$ 3.7} & \scriptsize{${\pm}$ 4.0} & \scriptsize{${\pm}$ 3.7} & \scriptsize{${\pm}$ 2.4} & \scriptsize{${\pm}$ 3.2} & \scriptsize{${\pm}$ 3.4} \\


& \multirow{1}{*}{$\ours$}  & \underline{84.7} & 74.7 & 61.3 & \textbf{84.4} & \underline{79.2} & \underline{68.5} & \textbf{68.7} & \underline{65.1} & \textbf{60.3} \\
& & \scriptsize{${\pm}$ 2.1 } & \scriptsize{${\pm}$ 3.4 } & \scriptsize{${\pm}$ 2.4 } & \scriptsize{${\pm}$ 1.3 } & \scriptsize{${\pm}$ 2.3 } & \scriptsize{${\pm}$ 3.1 } & \scriptsize{${\pm}$ 1.8 } & \scriptsize{${\pm}$ 3.2 } & \scriptsize{${\pm}$ 3.8 } \\

& \multirow{1}{*}{$\ours$-SE}  & 84.6 & 74.6 & 62.5 & 81.8 & 76.9 & \textbf{69.8} & \underline{68.5} & 64.3 & \underline{59.7} \\
& & \scriptsize{${\pm}$ 2.2 } & \scriptsize{${\pm}$ 2.8 } & \scriptsize{${\pm}$ 2.6 } & \scriptsize{${\pm}$ 2.0 } & \scriptsize{${\pm}$ 2.8 } & \scriptsize{${\pm}$ 2.3 } & \scriptsize{${\pm}$ 2.0 } & \scriptsize{${\pm}$ 2.2 } & \scriptsize{${\pm}$ 3.9 } \\

\midrule


\parbox[t]{3mm}{\multirow{16}{*}{\rotatebox[origin=c]{90}{\textbf{Prototyping}}}} & \multirow{1}{*}{\textsc{Mean}}  & 79.6 & 70.8 & 61.1 & 76.9 & 74.8 & 65.5 & 62.5 & 58.7 & 54.3 \\
& & \scriptsize{${\pm}$ 4.0} & \scriptsize{${\pm}$ 2.0} & \scriptsize{${\pm}$ 3.4} & \scriptsize{${\pm}$ 2.2} & \scriptsize{${\pm}$ 2.7} & \scriptsize{${\pm}$ 2.9} & \scriptsize{${\pm}$ 2.8} & \scriptsize{${\pm}$ 3.9} & \scriptsize{${\pm}$ 4.2} \\

& \multirow{1}{*}{Intra}  & 79.9 & 71.6 & 60.3 & 77.7 & 74.7 & 66.8 & 63.3 & 59.4 & 54.2 \\
& & \scriptsize{${\pm}$ 2.8 } & \scriptsize{${\pm}$ 1.4 } & \scriptsize{${\pm}$ 2.4 } & \scriptsize{${\pm}$ 1.8 } & \scriptsize{${\pm}$ 2.1 } & \scriptsize{${\pm}$ 3.2 } & \scriptsize{${\pm}$ 2.7 } & \scriptsize{${\pm}$ 3.8 } & \scriptsize{${\pm}$ 4.4 } \\

& \multirow{1}{*}{HIPT$_{\text{CLS-4k}}$ \cite{chen2022scaling}}  & 69.9 & 64.9 & 60.7 & 56.4 & 56.2 & 53.3 & 60.7 & 57.0 & 55.0 \\
& & \scriptsize{${\pm}$ 2.9} & \scriptsize{${\pm}$ 2.1} & \scriptsize{${\pm}$ 3.3} & \scriptsize{${\pm}$2.8} & \scriptsize{${\pm}$4.1} & \scriptsize{${\pm}$2.3} & \scriptsize{${\pm}$ 2.0} & \scriptsize{${\pm}$ 2.6} & \scriptsize{${\pm}$ 2.7} \\

& \multirow{1}{*}{GigaSSL \cite{lazard2023giga}}  & 80.0 & 71.5 & 62.8 & -- & -- & -- & -- & -- & -- \\
& & \scriptsize{${\pm}$ 2.4} & \scriptsize{${\pm}$ 2.0} & \scriptsize{${\pm}$ 2.4} & -- & -- & -- & -- & -- & -- \\

& \multirow{1}{*}{GigaPath-(\textsc{Mean}) \cite{xu2024gigapath}}  & 71.7 & 66.1  & 56.4 & 69.7 & 68.3 & 61.6 & 62.0 & 58.8 & 53.8 \\
& & \scriptsize{5.6} & \scriptsize{3.8} & \scriptsize{2.6} & \scriptsize{${\pm}$ 3.2} & \scriptsize{${\pm}$ 5.1} & \scriptsize{${\pm}$ 3.7} & \scriptsize{${\pm}$ 2.7} & \scriptsize{${\pm}$ 2.8} & \scriptsize{${\pm}$ 3.8} \\

& \multirow{1}{*}{GigaPath \cite{xu2024gigapath}}  & 70.5 & 63.9  & 56.4 & 68.5 & 66.5 & 59.4 & 60.4 & 57.4 & 52.5 \\
& & \scriptsize{5.4} & \scriptsize{2.7} & \scriptsize{1.9} & \scriptsize{${\pm}$ 3.3} & \scriptsize{${\pm}$ 4.8} & \scriptsize{${\pm}$ 4.0} & \scriptsize{${\pm}$ 2.9} & \scriptsize{${\pm}$ 2.4} & \scriptsize{${\pm}$ 3.4} \\


& \multirow{1}{*}{$\ours$}  & \textbf{85.1} & \textbf{76.4} & \underline{62.6} & \underline{83.0} & \textbf{80.7} & \underline{68.5} & 68.2 & \textbf{65.7} & 57.1 \\
& & \scriptsize{${\pm}$ 1.4 } & \scriptsize{${\pm}$ 1.2 } & \scriptsize{${\pm}$ 2.9 } & \scriptsize{${\pm}$ 1.6 } & \scriptsize{${\pm}$ 1.8 } & \scriptsize{${\pm}$ 2.2 } & \scriptsize{${\pm}$ 2.7 } & \scriptsize{${\pm}$ 3.8 } & \scriptsize{${\pm}$ 3.2 } \\

& \multirow{1}{*}{$\ours$-SE}  & 83.3 & \underline{74.9} & \textbf{62.9} & 80.5 & 77.3 & \textbf{69.8} & 67.2 & 64.7 & 56.9 \\
& & \scriptsize{${\pm}$ 1.5 } & \scriptsize{${\pm}$ 1.1 } & \scriptsize{${\pm}$ 3.1 } & \scriptsize{${\pm}$ 1.6 } & \scriptsize{${\pm}$ 1.5 } & \scriptsize{${\pm}$ 2.0 } & \scriptsize{${\pm}$ 2.6 } & \scriptsize{${\pm}$ 3.2 } & \scriptsize{${\pm}$ 3.8 } \\

\bottomrule
\end{tabular}
}
\end{table*}

\begin{table*}
\centering
\caption{\textbf{IHC quantification.} We quantify the abundance of estrogen (ER) (N=962) and progesterone (PR) (N=1,071) receptor expression in 3-class and 6-class scenarios using IHC. We compare $\ours$ fine-tuned against $\ours$ architecture trained from scratch and \textsc{Mean}. Results using 5-fold cross-validation with $k$=25 examples per class and evaluated using macro-AUC. Best in \textbf{bold}, second best is \underline{underlined}. 
}
\label{tab:ihc_quant}
\scalebox{1.0}{
\begin{tabular}{l|cc|cc}
\toprule
Model/Data & \multicolumn{2}{c}{\textbf{ER} ($\uparrow$)} & \multicolumn{2}{c}{\textbf{PR} ($\uparrow$)} \\
& $\mathcal{C}=3$ & $\mathcal{C}=6$ & $\mathcal{C}=3$ & $\mathcal{C}=6$  \\
\midrule
\multirow{1}{*}{\textsc{Mean} (linear probe)} & 74.6 \scriptsize{$\pm$ 1.9} & 69.5 \scriptsize{$\pm$ 1.2} & 73.1 \scriptsize{$\pm$ 1.8} & 69.1 \scriptsize{$\pm$ 1.0} \\

\multirow{1}{*}{\textsc{ABMIL} (Random)} & \underline{82.1} \scriptsize{$\pm$ 2.0} & \underline{83.4} \scriptsize{$\pm$ 1.3} & \underline{83.8} \scriptsize{$\pm$ 1.4} & \underline{83.9} \scriptsize{$\pm$ 1.4} \\

\multirow{1}{*}{\textsc{ABMIL} (FineTune)} & \textbf{89.6} \scriptsize{$\pm$ 1.3} & \textbf{86.0} \scriptsize{$\pm$ 0.8} & \textbf{89.4} \scriptsize{$\pm$ 0.9} & \textbf{85.5} \scriptsize{$\pm$ 0.9} \\

\bottomrule
\end{tabular}
}
\end{table*}

\begin{table*}
\centering
\caption{\textbf{Survival outcome prediction in TCGA Breast.} Models are trained using site-stratified 5-fold cross-validation. Evaluation using Concordance index (c-index). Besides HIPT, GigaSSL and GigaPath, all models use the \emph{same} patch encoder. Best in \textbf{bold}, second best is \underline{underlined}.}
\label{tab:survival}
\scalebox{1.0}{
\begin{tabular}{ll|c}
\toprule
& Model/Data & \textbf{TCGA Breast} ($\uparrow$)\\
\midrule
\parbox[t]{3mm}{\multirow{4}{*}{\rotatebox[origin=c]{90}{\textbf{MIL}}}} 
& \multirow{1}{*}{ABMIL \cite{ilse2018attention}}  & 0.669 $\pm$ 0.073    \\

& \multirow{1}{*}{TransMIL \cite{shao2021transmil}}  & \underline{0.697} $\pm$ 0.046   \\

& \multirow{1}{*}{IB-MIL \cite{li2023task}}  & 0.612 $\pm$ 0.088    \\

& \multirow{1}{*}{ILRA \cite{xiang2022exploring}}  & 0.657 $\pm$ 0.067   \\

\midrule
\parbox[t]{3mm}{\multirow{8}{*}{\rotatebox[origin=c]{90}{\textbf{Slide level}}}} 
& \multirow{1}{*}{\textsc{Mean}}  & 0.687 $\pm$ 0.079   \\

& \multirow{1}{*}{\textsc{Intra}}  & 0.692 $\pm$ 0.069    \\

& \multirow{1}{*}{HIPT$_{\text{CLS-4k}}$ \cite{chen2022scaling}}  & 0.547 $\pm$ 0.078    \\

& \multirow{1}{*}{GigaSSL \cite{lazard2023giga}}  & 0.530 $\pm$ 0.038   \\

& \multirow{1}{*}{GigaPath-(\textsc{Mean}) \cite{xu2024gigapath}}  & 0.587 $\pm$ 0.091   \\

& \multirow{1}{*}{GigaPath \cite{xu2024gigapath}}  & 0.521 $\pm$ 0.083   \\

& \multirow{1}{*}{$\ours$}  & \textbf{0.715} $\pm$ 0.041  \\

& \multirow{1}{*}{$\oursSE$}  & 0.696 $\pm$ 0.073  \\



\bottomrule
\end{tabular}
}
\end{table*}


\begin{table*}
\centering
\caption{\textbf{Molecular subtyping from H\&E.} Detection of HER2 and KI67 status (binary) from H\&E in AIDPATH and BCNB datasets. Results of $\ours$ and $\oursSE$ obtained using linear probing. ``SL" stands for slide level. Results using 5-fold stratified cross-validation evaluated using macro-AUC. Best in \textbf{bold}, second best is \underline{underlined}.}
\label{tab:aidpath_bcnb}
\scalebox{1.0}{
\begin{tabular}{ll|c|cc}
\toprule
& Model/Data & \textbf{HER2} ($\uparrow$) & \multicolumn{2}{c}{\textbf{KI67} ($\uparrow$)} \\
& & AIDPATH & AIDPATH & BCNB  \\
\midrule
\parbox[t]{3mm}{\multirow{4}{*}{\rotatebox[origin=c]{90}{\textbf{MIL}}}} 
& \multirow{1}{*}{ABMIL \cite{ilse2018attention}} & 81.1 \scriptsize{$\pm$ 8.9} & \underline{89.2} \scriptsize{$\pm$ 7.7} & \underline{81.9} \scriptsize{$\pm$ 3.7} \\

& \multirow{1}{*}{TransMIL \cite{shao2021transmil}} & 46.4 \scriptsize{$\pm$ 10.7} & 65.1 \scriptsize{$\pm$ 19.9} & 74.9 \scriptsize{$\pm$ 10.1} \\

& \multirow{1}{*}{IB-MIL \cite{li2023task}} & 73.2 \scriptsize{$\pm$ 11.1} & 87.7 \scriptsize{$\pm$ 6.1} & 81.6 \scriptsize{$\pm$ 3.5} \\

& \multirow{1}{*}{ILRA \cite{xiang2022exploring}} & 76.1 \scriptsize{$\pm$ 7.8} & 84.9 \scriptsize{$\pm$ 4.9} & 78.8 \scriptsize{$\pm$ 3.6} \\

\midrule

\parbox[t]{3mm}{\multirow{4}{*}{\rotatebox[origin=c]{90}{\textbf{SL}}}} 
& \multirow{1}{*}{\textsc{Mean}}  & 77.4 \scriptsize{$\pm$ 20.5} & 80.2 \scriptsize {$\pm$ 2.8} & 79.6  \scriptsize {$\pm$ 4.0} \\

& \multirow{1}{*}{\textsc{Intra}}  & \underline{85.8}  \scriptsize {$\pm$ 17.4} & 80.2  \scriptsize {$\pm$ 5.9} & 80.9  \scriptsize {$\pm$ 3.6} \\

& $\ours$  & 81.5  \scriptsize {$\pm$ 9.9} & \textbf{91.3}  \scriptsize {$\pm$ 4.9} & 81.4  \scriptsize {$\pm$ 4.2} \\ 

& $\oursSE$  & \textbf{92.5} \scriptsize{$\pm$ 7.2} & 83.0 \scriptsize{$\pm$ 8.6} & \textbf{82.0} \scriptsize{$\pm$ 3.6} \\



\bottomrule
\end{tabular}
}
\end{table*}

\clearpage
\section{Additional kidney results}
\label{supp:add_kidney}

\begin{table*}
\centering
\caption{\textbf{Kidney rejection tasks.} Linear probe and prototyping for $k=50$ reported. HIPT and GigaSSL are not available for non-cancer datasets. Best in \textbf{bold}, second best is \underline{underlined}.}
\label{tab:kidney}
\scalebox{1.0}{
\begin{tabular}{ll|cc|cc}
\toprule
& Model/Data & \multicolumn{2}{c}{\textbf{IFTA} ($\uparrow$)} & \multicolumn{2}{c}{\textbf{AMR} ($\uparrow$)} \\
& & \textbf{Slide} & \textbf{Patient} & \textbf{Slide} & \textbf{Patient} \\
\midrule
\parbox[t]{3mm}{\multirow{8}{*}{\rotatebox[origin=c]{90}{\textbf{MIL}}}} 
& \multirow{1}{*}{ABMIL \cite{ilse2018attention}} & 74.5 & 78.2 & 69.6 & 71.2 \\
& & \scriptsize{$\pm$ 1.9} & \scriptsize{$\pm$ 4.9} & \scriptsize{$\pm$ 3.7} & \scriptsize{$\pm$ 5.8} \\

& \multirow{1}{*}{TransMIL \cite{shao2021transmil}} & 57.5 & 60.4 & 55.9 & 55.4 \\
& & \scriptsize{$\pm$ 3.7} & \scriptsize{$\pm$ 6.6} & \scriptsize{$\pm$ 5.4} & \scriptsize{$\pm$ 10.5} \\

& \multirow{1}{*}{IB-MIL \cite{li2023task}} & 73.0 & 80.0 & 67.4 & 69.6  \\
& & \scriptsize{$\pm$ 3.7} & \scriptsize{$\pm$ 5.1} & \scriptsize{$\pm$ 4.3} & \scriptsize{$\pm$ 8.5}  \\

& \multirow{1}{*}{ILRA \cite{xiang2022exploring}} & 70.9 & 77.5 & 62.4 & 63.5 \\
& & \scriptsize{$\pm$ 5.0} & \scriptsize{$\pm$ 4.5} & \scriptsize{$\pm$ 6.8} & \scriptsize{$\pm$ 10.4} \\

\midrule


\parbox[t]{3mm}{\multirow{8}{*}{\rotatebox[origin=c]{90}{\textbf{Linear probe}}}} 
& \multirow{1}{*}{\textsc{Mean}} & 73.6 & 79.3 & 67.8  & 70.9 \\
& & \scriptsize{$\pm$ 2.6} & \scriptsize{$\pm$ 2.1} & \scriptsize{$\pm$ 3.2} & \scriptsize{$\pm$ 3.2} \\

& \multirow{1}{*}{\textsc{Intra}} & 74.5 & 80.2 & 67.9 & 72.0 \\
& & \scriptsize{$\pm$ 2.2} & \scriptsize{$\pm$ 1.8} & \scriptsize{$\pm$ 3.1} & \scriptsize{$\pm$ 3.2}  \\

& \multirow{1}{*}{$\ours$} & \underline{75.3} & \underline{80.9} & \textbf{71.2} & \underline{73.8} \\
& & \scriptsize{$\pm$ 2.5} & \scriptsize{$\pm$ 2.3} & \scriptsize{2.9} & \scriptsize{$\pm$ 3.2} \\

& \multirow{1}{*}{$\oursSE$} & \textbf{76.1} & \textbf{82.4} & \underline{70.0} & \textbf{74.2} \\
& & \scriptsize{$\pm$ 2.0} & \scriptsize{$\pm$ 1.8} & \scriptsize{$\pm$ 2.9} & \scriptsize{$\pm$ 3.5} \\

\midrule


\parbox[t]{3mm}{\multirow{8}{*}{\rotatebox[origin=c]{90}{\textbf{Prototyping}}}} 
& \multirow{1}{*}{\textsc{Mean}} & 70.2 & 75.0 & 63.8 & 67.5 \\
& & \scriptsize{$\pm$ 2.4} & \scriptsize{$\pm$ 2.8} & \scriptsize{$\pm$ 6.0} & \scriptsize{$\pm$ 8.1} \\

& \multirow{1}{*}{\textsc{Intra}} & 70.6 & 75.7 & 63.4 & 66.5 \\
& & \scriptsize{$\pm$ 2.5} & \scriptsize{$\pm$ 3.2} & \scriptsize{$\pm$ 4.3} & \scriptsize{$\pm$ 5.4} \\

& \multirow{1}{*}{$\ours$}  & 72.1 & 77.0 & 66.7 & 71.2 \\
& & \scriptsize{$\pm$ 2.4} & \scriptsize{$\pm$ 3.0} & \scriptsize{$\pm$ 3.7} & \scriptsize{$\pm$ 4.3}\\

& \multirow{1}{*}{$\oursSE$} & 73.1 & 79.8 & 65.5 & 70.4 \\
& & \scriptsize{$\pm$ 2.4} & \scriptsize{$\pm$ 2.0} & \scriptsize{$\pm$ 4.1} & \scriptsize{$\pm$ 5.4} \\

\bottomrule
\end{tabular}
}
\end{table*}

\clearpage
\section{Additional ablations}
\label{supp:add_ablations}

\begin{table*}[h]
\centering
\caption{\textbf{Ablation study of $\ours$ feature extractor.} Survival was evaluated using c-index and site-stratified 5-fold cross-validation. Subtyping and molecular status prediction were evaluated using macro-AUC and prototyping evaluation ($k$=25) repeated five times with fixed seed across baselines. Standard deviation reported over the 5 runs. $\ours$ refers to pretraining on breast cancer using \textsc{InfoNCE + GOT} without stain encoding. CONCH is the patch encoder of the Vision+Language model proposed in~\cite{lu2024towards}. Best in \textbf{bold}, second best is \underline{underlined}.}
\label{tab:ablation_feats}
\scalebox{1.0}{
\begin{tabular}{l|ccccc}
\toprule
Model/Data & \textbf{TCGA} & \textbf{BWH} & \textbf{TCGA} & \textbf{BCNB} & \textbf{Avg}\\
& \textbf{Survival} ($\uparrow$) & \textbf{Subtyping} ($\uparrow$) & \textbf{PR} ($\uparrow$) & \textbf{ER} ($\uparrow$) &\\
\midrule


\textsc{CTransPath+Mean} & 68.6 & 81.1 & 65.0 & 67.7 & 70.6 \\
& \scriptsize{$\pm$ 4.0} & \scriptsize{$\pm$ 3.9} & \scriptsize{$\pm$ 1.5} & \scriptsize{$\pm$ 2.1} &  \\

\textsc{CONCH+Mean} & 68.7 & 86.2 & 70.8 & 76.9 & 75.6 \\
& \scriptsize{$\pm$} 7.9 & \scriptsize{$\pm$ 7.9} & \scriptsize{$\pm$ 3.0} & \scriptsize{$\pm$ 2.0} &  \\

\midrule

\textsc{CTransPath}+$\ours$ & 65.4 & 83.1 & 66.9 & 68.6 & 71.0 \\
& \scriptsize{$\pm$ 6.5} & \scriptsize{$\pm$ 4.6} & \scriptsize{$\pm$ 1.6} & \scriptsize{$\pm$ 3.5} & \\

\textsc{CONCH}+$\ours$ & \textbf{71.5} & \textbf{94.9} & \textbf{76.4} & \textbf{83.0} & \textbf{81.5} \\
& \scriptsize{$\pm$ 4.1} & \scriptsize{$\pm$ 0.9} & \scriptsize{$\pm$ 1.9} & \scriptsize{$\pm$ 1.6} & \\
\bottomrule
\end{tabular}
}
\end{table*}

\begin{table*}[h]
\centering
\caption{\textbf{Ablation study of $\ours$ architecture.} Survival was evaluated using c-index and site-stratified 5-fold cross-validation. Subtyping and molecular status prediction were evaluated using macro-AUC and prototyping evaluation ($k$=25) repeated five times with fixed seed across baselines. Standard deviation reported over the 10 runs. ``MH'' refers to multi-head attention, ``SH'' to single-head attention, and ``SE'' to stain encoding. $\ours$ refers to pretraining on breast cancer using \textsc{InfoNCE + GOT}.
Best in \textbf{bold}, second best is \underline{underlined}.}
\label{tab:ablation_arch}
\scalebox{1.0}{
\begin{tabular}{l|ccccc}
\toprule
Model/Data & \textbf{TCGA} & \textbf{BWH} & \textbf{TCGA} & \textbf{BCNB} & \textbf{Avg}\\
& \textbf{Survival} ($\uparrow$) & \textbf{Subtyping} ($\uparrow$)& \textbf{PR} ($\uparrow$)& \textbf{ER} ($\uparrow$)&\\

\midrule

$\ours$-SH & 70.1 & 91.8 & 75.0 & 80.5 & 79.4 \\
& \scriptsize{$\pm$} 7.1 & \scriptsize{$\pm$} 1.7 & \scriptsize{$\pm$ 1.4} & \scriptsize{$\pm$ 2.5} &  \\
$\ours$ w. TransMIL & 55.7 & 90.8 & 75.6 & 82.1 & 76.5 \\
 & \scriptsize{$\pm$ 7.8} & \scriptsize{$\pm$ 1.7} & \scriptsize{$\pm$ 1.3} & \scriptsize{$\pm$ 1.3} &  \\
 $\ours$ w. SE & 69.6 & \textbf{95.8} & 74.9 & 80.5 & 80.2 \\
 & \scriptsize{$\pm$ 7.3} & \scriptsize{$\pm$ 0.8} & \scriptsize{$\pm$ 1.1} & \scriptsize{$\pm$ 1.6} &  \\
 $\ours$-MH & \textbf{71.5}  & 94.9 & \textbf{76.4 }& \textbf{83.0} & 81.5 \\
& \scriptsize{$\pm$} 4.1 & \scriptsize{$\pm$ 0.9} & \scriptsize{$\pm$ 1.2} & \scriptsize{$\pm$ 1.6} & \\

\bottomrule
\end{tabular}
}
\end{table*}

\clearpage
\section{Additional interpretability examples}
\label{supp:add_interpret}

\begin{figure*}[!ht]
   \centering
   \includegraphics[width=\linewidth]{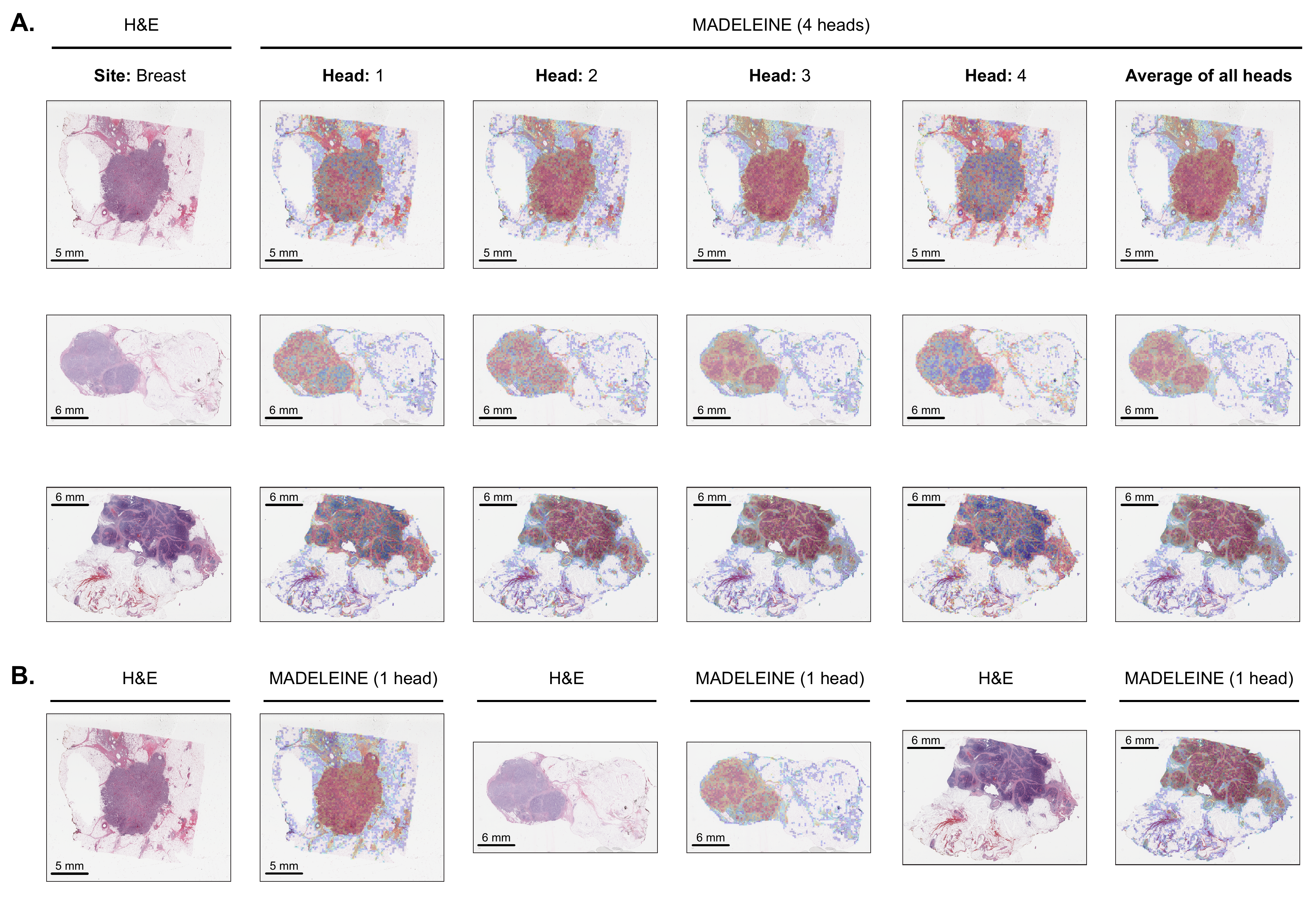}
   \caption{\textbf{Additional heatmap examples obtained with $\ours$} 
   \textbf{A.} Attention weights of multi-headed (frozen) ABMIL slide encoder pretrained with $\ours$ overlaid on three randomly chosen samples for TCGA Breast cohort. We show all heads and the average of heads. \textbf{B.} Attention weights of a single head (frozen) ABMIL slide encoder pretrained with $\ours$ overlaid on three randomly chosen samples for TCGA Breast cohort. Multi-headed ABMIL trained with $\ours$ can focus on different morphologies, whereas single-headed ABMIL focuses only on tumor morphology. 
   }
   \label{fig:inter}
\end{figure*}

\clearpage
\section{Limitations}
\label{supp:limitations}

$\ours$ is a multimodal pre-training strategy for slide representation learning. It operates under the assumption that representation learning of H\&E images can be guided by other stains (immunohistochemistry and special stains). This premise is directly inspired by the standard practice in clinical settings, where H\&E staining is routinely performed as the \textit{gold standard} procedure, along with complementary stains. Though this approach is principled, we highlight some limitations of our study and this methodology more broadly. \\

\noindent\textbf{Data scaling} Clinical practice is complex and ever evolving. Every year, new IHC and special stains become available, some of which are integrated in the workflow and can be used on a case-by-case basis. In breast cancer, our study focuses on four IHC stains (the most common ones), whereas many more can be employed, such as Epidermal Growth Factor Receptor (EGFR), P53, and E-Cadherin. As each stain offers a different view of a biomarker, increasing the number of stains would make the training signal richer and the resulting representation potentially better. \\ 

\noindent\textbf{Lack of large public datasets} Acrobat is the only \textit{large-scale} public dataset with H\&E and IHC stains. Therefore, without relying on proprietary data, such method cannot be scaled to more stains and other types of cancer. While the NADT-Prostate~\cite{wilkinson2020nascent} cohort includes H\&E and IHC, it remains limited by its size; for example, 14/18 stains provided have less than 100 examples, preventing efficient pre-training in prostate adenocarcinoma. In addition, TCGA includes known limitations such as site-specific biases \cite{howard2021impact} and demographic biases \cite{vaidya2024demographic}. Despite these limitations, TCGA remains the largest public resource for cancer prognostication and survival analyses. \\

\noindent\textbf{Model scaling} $\ours$ is trained using a combination of a global objective using contrastive learning and a local objective using graph optimal transport (GOT). Using our current hardware (3$\times$ 3090 GPUs), we are limited by the maximum batch size for contrastive learning, even using efficient parallelization and bfloat16 quantization. In addition, computing GOT is computationally expensive, with significant memory requirements. Because of this constraint, we must use 256 patch embeddings (or tokens) per stain for computing GOT. Scaling to more tokens would allow finer-grained matching between stains. Local alignment through GOT also requires morphological overlap between tissue sections used for different stains.  


\end{document}